\documentclass[10pt,journal]{IEEEtran}
\usepackage{tikz}
\usepackage{ifthen}
\usepackage{enumitem}
\usepackage{placeins}
\usepackage[T1]{fontenc}
\usepackage[utf8]{inputenc}
\usepackage{amsmath,amssymb,bm}
\usepackage{graphicx}
\usepackage[percent]{overpic}
\usepackage{booktabs}
\usepackage{array}
\usepackage{footnote}
\usepackage{multirow}
\usepackage{array}
\usepackage[T1]{fontenc}% optional T1 font encoding
\usepackage{setspace}
\usepackage[cmintegrals]{newtxmath}
\usepackage{algorithm}
\usepackage{algorithmic}
\usepackage{placeins}
\usepackage{stfloats}
\usepackage{cite}
\usepackage{url}
\usepackage[table]{xcolor}
\usepackage{titlesec}
\makeatletter
\providecommand{\subparagraph}{%
  \@startsection{subparagraph}{5}{\parindent}{3.25ex plus 1ex minus .2ex}%
  {-1em}{\normalfont\normalsize\bfseries}%
}
\makeatother

\graphicspath{{figures/}}
\definecolor{proposedrow}{RGB}{232,241,250}
\newcommand{\Nset}{\mathcal{N}}
\newcommand{\Rsl}{\mathcal{R}^{\mathrm{SL}}}
\newcommand{\Ruu}{\mathcal{R}^{\mathrm{Uu}}}
\newcommand{\Psl}{\mathcal{P}^{\mathrm{SL}}}
\newcommand{\Gcal}{\mathcal{G}}
\newcommand{\Ecal}{\mathcal{E}}
\newcommand{\SL}{\mathrm{SL}}
\newcommand{\Uu}{\mathrm{Uu}}
\newcommand{\SINR}{\mathrm{SINR}}
\newcommand{\ind}{\mathbb{I}}
\newcommand{\pos}[1]{\left[#1\right]_{+}}

\setlist[itemize]{topsep=0pt,partopsep=0pt,itemsep=0pt,parsep=0pt}
\setlist[enumerate]{topsep=0pt,partopsep=0pt,itemsep=0pt,parsep=0pt}
\titlespacing{\section}{0pt}{0.35ex plus 0.1ex minus 0.1ex}{0.25ex plus 0.1ex}
\titlespacing{\subsection}{0pt}{0.25ex plus 0.1ex minus 0.1ex}{0.15ex plus 0.1ex}
\titleformat{\subsubsection}[runin]{\normalfont\normalsize\itshape}{\arabic{subsubsection})}{0.5em}{}[.]
\titlespacing{\subsubsection}{0pt}{0.2ex plus 0.1ex minus 0.1ex}{0.1ex plus 0.1ex}
\AtBeginDocument{%
  \setlength{\abovedisplayskip}{2pt plus 1pt minus 1pt}%
  \setlength{\belowdisplayskip}{2pt plus 1pt minus 1pt}%
  \setlength{\abovedisplayshortskip}{1pt plus 0.5pt minus 0.5pt}%
  \setlength{\belowdisplayshortskip}{1pt plus 0.5pt minus 0.5pt}%
  \setlength{\itemsep}{0pt plus 0ex}%
}

\newtheorem{theorem}{Theorem}

\begin{document}

\title{Physical-Risk-Aware Decentralized Agents\\
for Constrained Vehicle Communications}

\author{Haoran Peng,~\IEEEmembership{Member,~IEEE},
William Weijia Zheng,~\IEEEmembership{Graduate Student Member,~IEEE},
Hang Liu,~\IEEEmembership{Member,~IEEE},
and Ying-Jun Angela Zhang,~\IEEEmembership{Fellow,~IEEE}%
\thanks{Haoran Peng, William Weijia Zheng, and Ying-Jun Angela Zhang are with the
Department of Information Engineering, The Chinese University of Hong Kong,
Shatin, N.T., Hong Kong S.A.R. (e-mail:
\{hrpeng, zw023, yjzhang\}@ie.cuhk.edu.hk).}%
\thanks{Hang Liu is with the State Key Laboratory of Internet of Things for
Smart City and the Department of Electronic and Communication, University of
Macau, Macao S.A.R. (e-mail: hangliu@um.edu.mo).}%
}

\markboth{}{}

\maketitle

\begin{abstract}
Decentralized V2X resource control is challenging because each vehicle has only local observations, while reliability depends on network-wide interactions. We develop a physical-risk-aware spatio-temporal graph attention (PA-STGAT) framework for agentic V2X control. Each vehicle perceives its local interaction graph, retains short-term temporal context, and independently selects its PC5 resource and power. Network outcomes are used during centralized training to refine the shared policy. Transmitter-specific credits map reliability failures to individual vehicles. A dual-priced graph term further converts network violations into coordination signals for discrete resource selection, with higher priority on short time-to-collision links. 
Experiments show higher raw-policy reliability than the tested graph-based baselines. 
Under filtered execution, PA-STGAT meets the nominal empirical communication requirements across all five seeds, with tradeoffs among utility, delivery reliability, resource occupancy, and filter-induced repair.
Compared with GNN+MATD3, it requires fewer resource repairs but achieves lower utility and mean delivery reliability. Mechanism ablations show that removing dual-priced graph coordination causes resource-selection collapse. 
Density and mobility tests characterize the operating limits of empirical constraint satisfaction.
\end{abstract}


\section{Introduction}
\label{sec:introduction}

\IEEEPARstart{V}{ehicle}-to-everything (V2X) communication is a key enabler of connected and autonomous transportation, supporting time-critical information exchange for cooperative awareness, safety, and infrastructure access \cite{3gpp22186}. 
In New Radio V2X (NR-V2X), vehicles may autonomously select PC5 sidelink resources without centralized scheduling, while the Uu interface provides vehicle-to-network connectivity \cite{3gpp23287}. 
Such distributed autonomy creates a fundamental mismatch between local decision making and network-wide consequences \cite{3gpp38300}. 
Each vehicle acts on locally available and potentially incomplete observations, whereas its resource and power choices affect packet delivery and interference across multiple links, including those involving unobserved receivers or interferers. 
Mobility further changes both these interactions and their physical significance. A communication failure between vehicles on conflicting trajectories with a short time-to-collision can be far more consequential than one on a noncritical link. 
The problem is therefore not simply a sequence of isolated resource-allocation decisions, but a closed-loop distributed-control task that requires continuous local perception, temporal context, coordinated actions, and network-level feedback \cite{11667010}. 
These characteristics make V2X a natural application domain for Agentic AI \cite{11673126}.

Realizing such agentic control in V2X remains challenging. Each vehicle observes only a local and possibly stale neighborhood, while mobility, fading, and blockage continuously change the communication environment.
A policy based only on the current observation may therefore miss important temporal changes in interference and physical risk. 
Propagation and packet characteristics directly affect NR-V2X quality of service~\cite{reyhanoglu2023qos}, while large vehicles can further degrade V2V reliability through shadowing~\cite{nguyen2023shadowing}. 
Cooperative perception may also rely on both PC5 and Uu links~\cite{mondal2026fusion}. 
As a result, intended receivers, potential interferers, and route-conflicting vehicles form different and time-varying relations.
A spatio-temporal representation is therefore more suitable than a single network snapshot; dynamic-graph V2X congestion control provides a related example~\cite{chen2023gatstc}. 
The communication requirements also differ in their temporal scope. Packet delivery, channel load, and infrastructure service are network-level constraints that can be regulated over time, whereas failures on short time-to-collision (TTC) links require tighter per-epoch control.
These requirements must also coexist with hard action-feasibility and local resource-separation rules. 
Their coupling with partial observation and decentralized decision making makes the problem difficult, since each vehicle must infer the evolving local context while the full impact of its action is only reflected in network-level outcomes.

Existing learning-based studies mainly address this problem from three directions. 
First, CTDE-based multi-agent reinforcement learning (MARL) uses global information during training, while each agent acts on local observations during execution~\cite{10701015}. 
Second, graph-based models capture spatial interactions among neighboring vehicles~\cite{10697115,10632058}.
Recurrent models complement them by using observation histories to recover temporal context under partial observability~\cite{10431724}. 
Third, constrained RL introduces explicit costs and adaptive Lagrangian multipliers to regulate reliability, latency, and safety requirements~\cite{10431724,10907776}.
These methods improve coordination, local state representation, and constraint handling. 
However, an important gap remains in V2X resource control. 
Network-level reliability and physical-risk violations must still be mapped to individual vehicle decisions. 
This is especially difficult for discrete resource selection, where the learning signal must identify both the responsible transmitters and the resource conflicts that caused the violation.

\subsection{Related Work}
\label{sec:related_work}

At the intersection of Agentic AI and V2X, existing studies mainly address three issues: risk-aware communication, dynamic representation of local interactions, and decentralized resource control under network constraints.

\noindent\textbf{Communication reliability and physical risk in V2X.}
Communication failures in V2X can have different consequences depending on the driving context. Denial-of-service studies quantify how network degradation affects vehicle risk~\cite{petho2024cyberrisk}. Platoon resource-selection methods reduce hidden-node and mobility-induced packet collisions~\cite{wang2025pcars}. Other works jointly optimize authentication, computation, communication, and control~\cite{lin2026riskaware}, or reduce communication load for timely collision prediction through semantic V2X~\cite{onsu2026semantic}. These studies show the importance of risk-aware communication. However, they do not directly address how a vehicle-local resource policy should respond when physical risk, interference, and network-wide reliability change together.

\noindent\textbf{Dynamic graph representations for V2X.}
Dynamic graphs provide a natural way to represent time-varying interactions among vehicles. They have been used for V2X congestion control~\cite{chen2023gatstc}, semantic- and intent-aware allocation with federated GraphSAGE~\cite{singh2026gfedmap}, and heterogeneous temporal platoon control~\cite{huang2026htdrive}. These methods use graph structure to improve the actor or critic representation. The constraint signal, however, is usually constructed outside the graph. As a result, network failures are not directly assigned to the responsible transmitters, and reliability prices do not act on the current interaction topology.

\noindent\textbf{Learning-based and constrained V2X resource control.}
Learning-based methods have also been widely used for decentralized V2X decisions. Multi-agent deep Q-networks support distributed spectrum and power control~\cite{liang2019spectrum}. Federated MARL addresses hidden-node conflicts in NR-V2X Mode~2~\cite{saad2025federatednrmode2}, while platoon actor--critic methods balance freshness, delivery, and energy~\cite{zheng2026aoiplatoon}. Agentic vehicular edge caching extends autonomous decisions to content placement and resource allocation~\cite{dai2026agenticcaching}. Other studies examine density-aware reservation and resource-control tradeoffs~\cite{zhou2025resource,wang2026multiagentdrlv2xresource}. In many of these methods, reliability enters learning through a fixed reward coefficient. CMDP-based approaches instead model reliability as an explicit cost budget~\cite{singh2023learningcmdp}, and augmented-Lagrangian methods adapt the corresponding price~\cite{li2021alm}. For vehicle-local categorical actions, however, the remaining problem is how to convert that network-level price into a transmitter-specific and topology-dependent learning signal.

\subsection{Proposed Framework and Contributions}
To address this gap, we develop a physical-risk-aware spatio-temporal graph attention (PA-STGAT) framework for decentralized V2X resource control. Each vehicle builds recent one-hop ego graphs and uses a shared ST-GAT actor to select its PC5 resource and power. During training, centralized critics evaluate the resulting network outcomes. Transmitter-specific credits then assign delivery failures to individual vehicles. Augmented-Lagrangian (AL) prices adapt to the measured constraint violations. For discrete resource selection, the Packed delivery ratio (PDR) price also weights a differentiable same-resource conflict term on the current graph. Short-TTC links receive higher weights. This mechanism turns network-level reliability feedback into a topology-dependent coordination signal for local resource decisions. Training executes the raw actor proposals so that their actual consequences remain visible. At deployment, each vehicle uses only its local actor and a one-hop repair filter, without centralized critics.
The main technical contributions are summarized as follows:
\begin{itemize}

    \item \textbf{Physical-risk-aware decentralized modeling:}
    We formulate joint PC5 resource and power control as a partially observed constrained problem. Each vehicle builds a local multi-relation graph for communication, interference, and TTC-based risk. A short graph history provides temporal context for decentralized decisions. The formulation also separates long-term network constraints, low-TTC reliability requirements, and structural action constraints.

    \item \textbf{Metric-preserving transmitter credit:}
    We derive transmitter-specific PDR and low-TTC costs from measured link failures. Their average exactly recovers the corresponding network-level metrics. The resulting credits provide per-vehicle cost targets while preserving the original constraint budgets and reward.

    \item \textbf{Dual-priced graph coordination:}
    We introduce a differentiable same-resource conflict term on the current graph. Its weight adapts with the PDR dual price, and short-TTC links receive higher priority. This creates a direct gradient from network-level reliability violations to local resource probabilities. It is designed for discrete resource coordination, where critic feedback alone can be indirect.

\end{itemize}

The experiments evaluate these mechanisms separately.
Graph aggregation improves the utility--occupancy tradeoff, and dual-priced graph coordination provides the strongest mechanism evidence: removing it causes deterministic resource-selection collapse.
The independent gains from transmitter credit, risk relations, and temporal recurrence are not statistically resolved in the nominal raw-policy comparisons.
Raw and filtered evaluations distinguish learned coordination from execution-time repair, while density and mobility tests characterize the operating limits.

The remainder of this paper is organized as follows. Section~\ref{sec:system} presents the system model and constrained formulation; Section~\ref{sec:method} develops PA-STGAT, its training and deployment mechanisms, and the theoretical analysis of feasibility and near-optimality;  Section~\ref{sec:experiments} reports the experimental results; and Section~\ref{sec:conclusion} concludes the paper.

\begin{table}[!t]
\centering
\caption{Core notation.}
\label{tab:notation}
\footnotesize
\begin{tabular}{@{}p{0.34\columnwidth}p{0.58\columnwidth}@{}}
\toprule
Symbol & Meaning\\
\midrule
\(\Nset,\Rsl,\Ruu,\Psl\) &
Vehicles, PC5/Uu resources, and discrete PC5 powers\\
\(a_{n,t},\tilde a_{n,t},a^{\mathrm{dep}}_{n,t}\) &
Communication action, raw proposal, and filtered action\\
\(\rho^{\SL}_{n,t},P^{\SL}_{n,t},y_{n,r,t}\) &
Selected PC5 resource/power and binary resource indicator\\
\(M_{n,m,t},S_{n,m,t},Z_{n,m,t}\) &
Intended-link, successful-delivery, and low-TTC-link indicators\\
\(d_{\mathrm{comm}}\) & Intended PC5 communication range\\
\(q^{\Uu}_{n,t},P^{\Uu}_{n,t}\) &
Exogenous Uu grant and transmit power\\
\(\kappa_{\Uu\rightarrow\SL},\kappa_{\SL\rightarrow\Uu}\) &
Synthetic cross-link leakage coefficients\\
\(R^{\SL}_{n,m,t},R^{\Uu}_{n,t}\) & PC5 link rate and vehicle Uu rate\\
\(\Gamma^{\SL},D_{\max},R^{\Uu}_{\min}\) &
Decoding threshold, packet-delay budget, and minimum Uu rate\\
\(\tau_{\mathrm{ref}},\tau_{\mathrm{met}}\) &
Physical-risk-edge horizon and low-TTC metric threshold\\
\(b_n^{\mathrm{ax}},\tau^{\mathrm{int}}_{n,t},d_{\mathrm{zone}},
\Delta_{\mathrm{int}}\) &
Route-axis index, center-arrival time, conflict-zone radius, and arrival slack\\
\(u_{n,t},u_t\) & Per-vehicle and vehicle-average network utility\\
\(L_{\mathrm{pkt}}\) & Packet size\\
\(\mathcal L_{(\cdot)}\) & Learning-objective losses\\
\(\mathrm{PDR}_t,\mathrm{CBR}_t\) &
Aggregate delivery-success and resource-occupancy metrics\\
\(v^{\mathrm{cl}}_{n,m,t},\tau^{\mathrm{rad}}_{n,m,t}\) &
Radial closing speed and TTC surrogate\\
\(\mathrm{HD}_{n,m,t},\bm f^\ell_{i,j,t}\) &
Half-duplex conflict indicator and relation-\(\ell\) edge features\\
\(\varepsilon_0\) & Positive numerical stabilizer\\
\(\Gcal_t,\Gcal^{\mathrm{ego}}_{n,t}\) &
Training graph and vehicle \(n\)'s one-hop graph\\
\(\Ecal_t^{(\cdot)},\Ecal_t^{\mathrm{prot}}\) &
Relation-specific graph edges and conditionally resource-separated protected edges\\
\(\mathbb E\) & Expectation operator\\
\(\bm x_{i,t},\bm h_{i,t},\bm z_{i,t}^{\mathrm{temp}}\) &
Node feature, spatial embedding, and temporal embedding\\
\(Q^u_{\psi_u,n},\bm Q^c_{\psi_c,n}\) &
Per-vehicle utility estimate and four cost estimates\\
\(\bm c_{n,t},\bm c_t,\bm\epsilon\) &
Per-vehicle costs, their \(N\times4\) stack, and budgets\\
\(\bm\lambda,\zeta_{\mathrm{TTC}},\rho_g\) &
AL dual prices, instantaneous TTC coefficient, and graph-conflict coefficient\\
\(\Pi_n,\mathcal F_{n,t},\widetilde{\mathcal P}_{n,t}\) &
Local filter, its feasible candidate set, and observed proposal messages\\
\bottomrule
\end{tabular}
\end{table}

\section{System Model and Problem Formulation}
\label{sec:system}

The radio action affects several links at once, but each vehicle observes
only a subset of them.  We specify those links and their PC5/Uu consequences,
derive risk-dependent communication priority from vehicle motion, and express
the resulting local decision under network budgets as a partially observed
CMDP.  Vehicle motion is exogenous; the optimized variables belong to the
radio layer.  Table~\ref{tab:notation} summarizes the notation.

\subsection{Vehicle-Communication Scenario and Time Scale}
\label{subsec:scenario}

\begin{figure}[t]
    \centering
    \begin{overpic}[width=0.95\columnwidth]{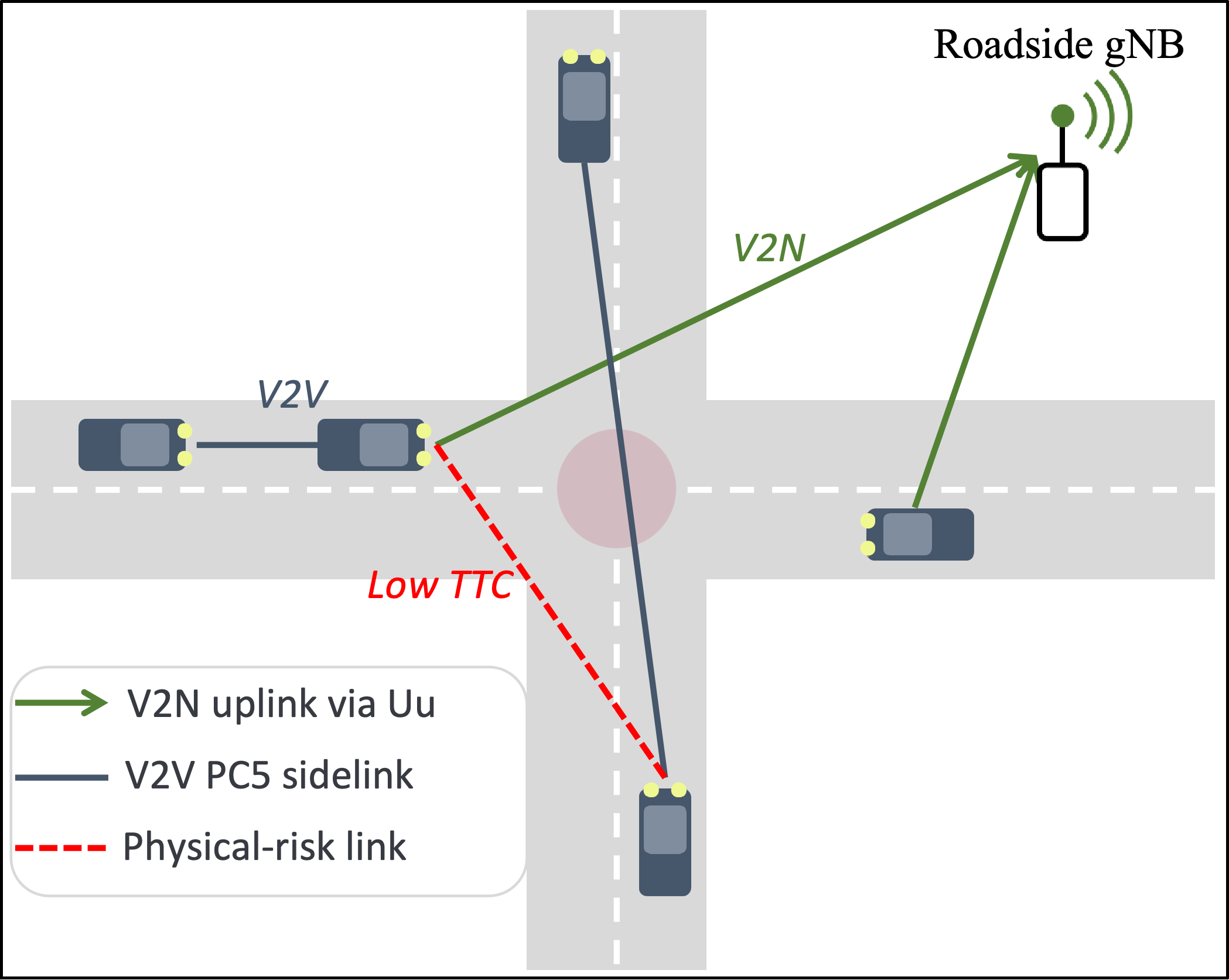}
    \end{overpic}
    \caption{Intersection scenario. Vehicles choose PC5 resource--power
    pairs for periodic V2V packets, and exogenously granted V2N uplinks
    terminate at the roadside gNB.  This gNB path is the modeled
    infrastructure link.  A TTC-derived relation marks the route-conflicting
    pair for elevated communication priority.}
    \label{fig:scenario}
    \vspace{-12pt}
\end{figure}

Figure~\ref{fig:scenario} shows the urban-intersection scenario.  The system
comprises the vehicle set
\(\Nset=\{1,\ldots,N\}\), one roadside next-generation NodeB (gNB), and two
coexisting communication paths.
Vehicles broadcast periodic V2V packets over the PC5 sidelink and use
exogenous gNB-scheduled Uu uplinks for V2N traffic.
The gNB supplies Uu grants but neither chooses vehicle-side PC5 actions nor
runs the deployed actor or its filter.
This architecture follows the PC5/Uu roles and autonomous sidelink selection
specified for NR-V2X~\cite{3gpp22186,3gpp23287,3gpp38300}.  The mobility and
channel assumptions follow V2X evaluation studies
~\cite{3gpp37885,3gpp38901}; the resulting system-level model abstracts the
detailed physical-layer (PHY) procedures of a particular 3GPP profile.

The controller uses \(\Delta t=10\) ms epochs as a system-level decision
interval.  Each epoch consists of local observation, one radio action, link
evaluation, and a mobility update
\begin{equation}
    \bm p_{n,t+1}=\bm p_{n,t}+\bm v_{n,t}\Delta t,
    \label{eq:motion}
\end{equation}
followed by reconstruction of every graph relation.  The environment generates
the trajectories and velocities.  The policy changes communication resources
and power; acceleration, braking, steering, and lane changes remain outside
its action space.

\subsection{Communication Action and Link Model}
\label{subsec:radio}

Each vehicle controls exactly two quantities: one PC5 resource and one
discrete transmit power,
\begin{equation}
    a_{n,t}=\left(\rho^{\SL}_{n,t},P^{\SL}_{n,t}\right),\quad
    \rho^{\SL}_{n,t}\in\Rsl,\quad
    P^{\SL}_{n,t}\in\Psl,
    \label{eq:action}
\end{equation}
where \(\SL\) and \(\Uu\) denote sidelink and Uu, and calligraphic symbols
denote their resource or power sets.  We use \(|\Rsl|=14\) and
\(\Psl=\{0.05,0.10,0.20\}\) W.  Uu grants and powers are generated
exogenously and provide interference and service context.  The actor output is
decoded into a valid proposal \(\tilde a_{n,t}\).  During filtered execution,
the local map \(\Pi_n\) may revise that proposal using the ego graph and the
observed proposal messages \(\widetilde{\mathcal P}_{n,t}\):
\(a^{\mathrm{dep}}_{n,t}=\Pi_n(\tilde a_{n,t};
\Gcal^{\mathrm{ego}}_{n,t},\widetilde{\mathcal P}_{n,t})\).  Raw evaluation
applies \(\tilde a_{n,t}\); filtered evaluation applies
\(a^{\mathrm{dep}}_{n,t}\).

One PC5 action has two network consequences.  Its resource and power determine
the desired V2V signal and co-channel interference, and spectral overlap also
couples it to the exogenous Uu service.  We model the PC5 consequence first.
Let \(g^{\SL}_{n,m,t}\) be the instantaneous channel
power gain from PC5
transmitter \(n\) to receiver \(m\), including distance-dependent path
loss, shadowing, and Rayleigh fading.  Let \(q^{\Uu}_{j,t}\) and
\(P^{\Uu}_{j,t}\) denote the exogenous Uu grant
\((q^{\Uu}_{j,t}\in\Ruu)\) and power, and let
\(\operatorname{ov}(r,q)=\ind\{r\bmod|\Ruu|=q\}\) be the simulator's resource-overlap
map.  The sidelink signal-to-interference-plus-noise ratio (SINR) and its
two interference terms are
\begin{align}
\SINR^{\SL}_{n,m,t}
&=
\frac{P^{\SL}_{n,t}g^{\SL}_{n,m,t}}
{\sigma^2+I^{\SL}_{n,m,t}+I^{\Uu\rightarrow\SL}_{n,m,t}},
\label{eq:sl_sinr}
\\
I^{\SL}_{n,m,t}
&=\sum_{i\notin\{n,m\}}
\ind\{\rho^{\SL}_{i,t}=\rho^{\SL}_{n,t}\}
P^{\SL}_{i,t}g^{\SL}_{i,m,t},\nonumber\\
I^{\Uu\rightarrow\SL}_{n,m,t}
&=\kappa_{\Uu\rightarrow\SL}\sum_j
\operatorname{ov}(\rho^{\SL}_{n,t},q^{\Uu}_{j,t})
P^{\Uu}_{j,t}g^{\Uu\rightarrow\SL}_{j,m,t},
\nonumber
\end{align}
where \(\sigma^2\) is noise power and
\(\kappa_{\Uu\rightarrow\SL}\) is the Uu-to-PC5 leakage coefficient.
If receiver \(m\) transmits on the same PC5 resource, half-duplex
operation sets that packet's effective SINR to zero.  Specifically, define
the half-duplex (HD) conflict indicator
\begin{equation}
\mathrm{HD}_{n,m,t}
=\ind\{\rho^{\SL}_{m,t}=\rho^{\SL}_{n,t}\},
\label{eq:half_duplex}
\end{equation}
where \(\mathrm{HD}_{n,m,t}=1\) denotes a half-duplex conflict on link
\((n,m)\).  With per-resource bandwidth \(B^{\SL}\), packet size
\(L_{\mathrm{pkt}}\), packet delay budget \(D_{\max}\), and decoding
threshold \(\Gamma^{\SL}\),
\begin{align}
R^{\SL}_{n,m,t}&=B^{\SL}\log_2(1+\SINR^{\SL}_{n,m,t}),\\
D^{\SL}_{n,m,t}&=\frac{L_{\mathrm{pkt}}}{R^{\SL}_{n,m,t}}.
\end{align}
The success indicator for an intended link is
\begin{equation}
\begin{split}
S_{n,m,t}=\ind\{&
\SINR^{\SL}_{n,m,t}\geq\Gamma^{\SL},\
D^{\SL}_{n,m,t}\leq D_{\max},\\[-2pt]
&\mathrm{HD}_{n,m,t}=0\}.
\end{split}
\label{eq:success}
\end{equation}
The reciprocal consequence is PC5-to-Uu leakage at the gNB.
Let \(b\) index the gNB and \(g^{\Uu}_{n,b,t}(q)\) be the gain from vehicle
\(n\) to that gNB on Uu resource \(q\).  The modeled Uu SINR and rate
are
\begin{align}
\SINR^{\Uu}_{n,t}
&=\frac{P^{\Uu}_{n,t}g^{\Uu}_{n,b,t}(q^{\Uu}_{n,t})}
{\sigma^2+I^{\Uu}_{n,t}+I^{\SL\rightarrow\Uu}_{n,t}},
\label{eq:uu_sinr}\\
I^{\Uu}_{n,t}
&=\sum_{i\ne n}\ind\{q^{\Uu}_{i,t}=q^{\Uu}_{n,t}\}
P^{\Uu}_{i,t}g^{\Uu}_{i,b,t}(q^{\Uu}_{n,t}),\nonumber\\
I^{\SL\rightarrow\Uu}_{n,t}
&=\kappa_{\SL\rightarrow\Uu}\sum_j
\operatorname{ov}(\rho^{\SL}_{j,t},q^{\Uu}_{n,t})
P^{\SL}_{j,t}g^{\Uu}_{j,b,t}(q^{\Uu}_{n,t}),\nonumber\\
R^{\Uu}_{n,t}
&=B^{\Uu}\log_2(1+\SINR^{\Uu}_{n,t}).
\label{eq:uu_rate}
\end{align}
The overlap map and leakage coefficients create action-dependent cross-link
coupling at the system level.  Their values are synthetic abstraction
parameters.  Protocol-level prediction would additionally require calibrated
spectral masks, numerology, and receiver procedures.

The learner needs one metric for successful delivery and another for spectrum
occupancy. Both are evaluated over the links that should receive a packet.
Let \(M_{n,m,t}\in\{0,1\}\) indicate such a link; a vehicle is an intended
receiver when it is within
communication range or belongs to a route-conflicting pair inside the
physical-risk horizon:
\begin{equation}
M_{n,m,t}
=\ind\{n\ne m\}
\ind\!\left\{d_{n,m,t}\leq d_{\mathrm{comm}}
\ \lor\ \tau^{\mathrm{rad}}_{n,m,t}\leq\tau_{\mathrm{ref}}\right\},
\label{eq:intended_link}
\end{equation}
where \(d_{\mathrm{comm}}=120\) m, and
\(\tau^{\mathrm{rad}}_{n,m,t}\) and \(\tau_{\mathrm{ref}}\) are defined in
Section~\ref{subsec:risk}.  The network-level PDR and CBR are
\begin{align}
\mathrm{PDR}_{t}
&=1-\frac{\sum_{n,m}M_{n,m,t}(1-S_{n,m,t})}
{\max(\sum_{n,m}M_{n,m,t},1)},\label{eq:pdr}\\
\mathrm{CBR}_{t}
&=\frac{1}{|\Rsl|}\sum_{r\in\Rsl}
\ind\!\left\{\sum_n\ind\{\rho^{\SL}_{n,t}=r\}>0\right\}.
\label{eq:cbr}
\end{align}
Here CBR denotes the occupied fraction of the abstract resource pool.  It is a
system-level load metric; calibrated 3GPP PHY sensing would require a more
detailed measurement model.  PDR and CBR summarize network delivery and occupancy; the next
subsection adds the pairwise priority needed to distinguish urgent links.

\subsection{Physical-Risk Sensing and Communication Priority}
\label{subsec:risk}

PDR gives every intended packet the same weight.  We introduce a separate
priority signal for pairs whose relative motion indicates a near-term route
conflict.  The construction first removes geometrically irrelevant pairs and
then computes TTC for the remaining closing pairs.  This two-stage test avoids
prioritizing nearby vehicles that are moving apart or following unrelated
routes.  For vehicles \((n,m)\), define relative position and velocity as
\(\Delta\bm p_{n,m,t}=\bm p_{n,t}-\bm p_{m,t}\) and
\(\Delta\bm v_{n,m,t}=\bm v_{n,t}-\bm v_{m,t}\), with distance
\(d_{n,m,t}=\|\Delta\bm p_{n,m,t}\|_2\).
The radial closing speed is the negative derivative of distance,
\begin{equation}
v^{\mathrm{cl}}_{n,m,t}
=-\frac{\Delta\bm p_{n,m,t}^{\mathsf T}\Delta\bm v_{n,m,t}}
{\max(d_{n,m,t},\varepsilon_0)},
\label{eq:closing_speed}
\end{equation}
where \(\varepsilon_0>0\) prevents division by zero.  A positive
\(v^{\mathrm{cl}}\) indicates decreasing separation; tangential motion leaves
the distance unchanged to first order.
Let \(b_n^{\mathrm{ax}}\in\{0,1\}\) identify the horizontal or vertical route
axis.  With the intersection center as the coordinate origin, the predicted
center-arrival time is
\begin{equation}
\tau^{\mathrm{int}}_{n,t}=
\begin{cases}
-\bm p_{n,t}^{\mathsf T}\bm v_{n,t}/\|\bm v_{n,t}\|_2^2,
&\bm p_{n,t}^{\mathsf T}\bm v_{n,t}<0,\\
+\infty,&\text{otherwise}.
\end{cases}
\label{eq:intersection_arrival}
\end{equation}
For \(n\neq m\), the implemented route-conflict gate is
\begin{align}
\chi^{\mathrm{same}}_{n,m,t}
&=\ind\!\left\{b_n^{\mathrm{ax}}=b_m^{\mathrm{ax}},\
d_{n,m,t}\leq d_{\mathrm{comm}},\right.\nonumber\\
&\hspace{7.3em}\left.v^{\mathrm{cl}}_{n,m,t}>0\right\},\nonumber\\
\chi^{\mathrm{cross}}_{n,m,t}
&=\ind\!\left\{b_n^{\mathrm{ax}}\neq b_m^{\mathrm{ax}},\
\max\{\tau^{\mathrm{int}}_{n,t},\tau^{\mathrm{int}}_{m,t}\}<+\infty,\right.\nonumber\\
&\hspace{3.1em}\left.
|\tau^{\mathrm{int}}_{n,t}-\tau^{\mathrm{int}}_{m,t}|
\leq\tau_{\mathrm{ref}},\right.\nonumber\\
&\hspace{3.1em}\left.
\min\{\tau^{\mathrm{int}}_{n,t},\tau^{\mathrm{int}}_{m,t}\}
\leq\tau_{\mathrm{ref}}+\Delta_{\mathrm{int}}\right\},\nonumber\\
\chi^{\mathrm{near}}_{n,m,t}
&=\ind\!\left\{b_n^{\mathrm{ax}}\neq b_m^{\mathrm{ax}},\
v^{\mathrm{cl}}_{n,m,t}>0,\right.\nonumber\\
&\hspace{3.1em}\left.
\min\{\|\bm p_{n,t}\|_2,\|\bm p_{m,t}\|_2\}
\leq d_{\mathrm{zone}}\right\},\nonumber\\
\chi_{n,m,t}
&=\max_{q\in\{\mathrm{same},\mathrm{cross},\mathrm{near}\}}
\chi^{q}_{n,m,t}.
\label{eq:route_gate}
\end{align}
We use \(d_{\mathrm{zone}}=45\) m and
\(\Delta_{\mathrm{int}}=1\) s.  The three branches correspond to rear-end or
head-on closure along one axis, predicted arrival overlap across axes, and
cross-axis closure already near the intersection.  Once this gate identifies
a plausible conflict, radial closure determines the TTC surrogate
\begin{equation}
\tau^{\mathrm{rad}}_{n,m,t}=
\begin{cases}
d_{n,m,t}/v^{\mathrm{cl}}_{n,m,t},
&v^{\mathrm{cl}}_{n,m,t}>\varepsilon_0,\ \chi_{n,m,t}=1,\\
+\infty,&\text{otherwise}.
\end{cases}
\label{eq:ttc}
\end{equation}
Before graph encoding, \(\tau^{\mathrm{rad}}_{n,m,t}\) is clipped at the
risk-edge horizon \(\tau_{\mathrm{ref}}=3\) s.  Equation~\eqref{eq:ttc}
equals constant-velocity TTC for collinear motion.  For crossing routes, it
serves as a communication-priority indicator derived from relative motion.
Shorter TTC receives larger weight
\begin{equation}
w_{n,m,t}=1+\alpha_r
\min\!\left\{w_{\max},
\frac{\tau_{\mathrm{ref}}}
{\max(\tau^{\mathrm{rad}}_{n,m,t},\varepsilon_0)}\right\}.
\label{eq:risk_weight}
\end{equation}
The 3-s horizon determines which risk edges enter the graph, and
\(w_{n,m,t}\) exposes their urgency to the learned attention mechanism.  A
stricter threshold, \(\tau_{\mathrm{met}}=1.5\) s, selects the subset used by
the low-TTC cost and constraint.  The wider graph horizon therefore gives the
actor advance context before a link enters the tighter metric.  Defining
\(Z_{n,m,t}=M_{n,m,t}
\ind\{\tau^{\mathrm{rad}}_{n,m,t}\leq\tau_{\mathrm{met}}\}\), its
communication-failure rate is
\begin{equation}
p^{\mathrm{TTC}}_{\mathrm{fail},t}
=\frac{\sum_{n,m}Z_{n,m,t}(1-S_{n,m,t})}
{\max(\sum_{n,m}Z_{n,m,t},1)}.
\label{eq:ttc_failure}
\end{equation}
Equation~\eqref{eq:ttc_failure} is therefore the delivery-failure rate over
the links assigned the highest radio priority.

Binary success hides how far a failed link lies from its SINR or delay target.
We retain that severity in a diagnostic used to interpret the
experiments:
\begin{align}
\delta^{\SINR}_{n,m,t}
&=\frac{\pos{\Gamma^{\SL}-\SINR^{\SL}_{n,m,t}}}{\Gamma^{\SL}},
\nonumber\\
\delta^{D}_{n,m,t}
&=\min\!\left\{d_{\mathrm{cap}},
\frac{\pos{D^{\SL}_{n,m,t}-D_{\max}}}{D_{\max}}\right\},\nonumber\\
H^{\mathrm{diag}}_{n,t}
&=\frac{\sum_m M_{n,m,t}w_{n,m,t}
(\delta^{\SINR}_{n,m,t}+\delta^D_{n,m,t}
+\mathrm{HD}_{n,m,t})}
{\max(\sum_m M_{n,m,t},1)},\nonumber\\
H_t
&=\frac{1}{N}\sum_n H^{\mathrm{diag}}_{n,t}.
\label{eq:hard_cost}
\end{align}
The delay gap equals \(d_{\mathrm{cap}}\) when the rate is zero.  The CMDP
budgets use the binary metrics above; \(H_t\) is reported as a TTC-weighted
diagnostic that explains the severity of their violations.

\subsection{Observation, State, and CMDP}
\label{subsec:cmdp}

The preceding quantities now define the information and objective of the
controller.  Centralized training uses a global state \(s_t\) containing
vehicle kinematics, route/conflict
attributes, intended-link indicators, recent resource-allocation and queue
context, current exogenous Uu grants, and the dynamic graph defined in
Section~\ref{sec:method}.  At execution, vehicle \(n\) receives its ego history
\(o_{n,t}\): its own row, one-hop neighbor rows, normalized message age,
observed/missing masks, and incident relation edges.  A common policy
\(\pi_\theta(a_{n,t}\mid o_{n,t-K+1:t})\) is executed separately by all
homogeneous vehicles, where \(K\) is the graph-history length and
\(\theta\) denotes the shared actor parameters.

The objective rewards Uu service and represents V2V reliability and load as
separate costs.  This separation allows AL to adapt each constraint price from
its own violation.  The utility is
\begin{align}
u_{n,t}&=\frac{R^{\Uu}_{n,t}}{R_0},\qquad
R_0=1~\mathrm{Mbit/s},\nonumber\\
u_t&=\frac{1}{N}\sum_{n=1}^{N}u_{n,t}.
\label{eq:utility}
\end{align}
The four cost components measure PC5 packet failure, low-TTC failure, CBR, and
the signed normalized Uu-rate residual.  For horizon \(T\), discount factor
\(\gamma\in(0,1)\), and deployment-filter collection
\(\Pi=\{\Pi_n\}_{n\in\Nset}\), the design problem is
\begingroup
\interdisplaylinepenalty=2500
\begin{align}
(\mathrm P):\quad
\max_{\theta}\quad&
\mathbb E_{\pi_\theta,\Pi}\!
\left[\sum_{t=1}^{T}\gamma^{t-1}u_t\right]
\label{eq:problem}\\
\mathrm{s.t.}\quad
&\frac{1}{T}\sum_t\mathbb E[1-\mathrm{PDR}_{t}]
\leq0.10,
\tag{C1-L}\label{eq:c1}\\
&\mathbb E[p^{\mathrm{TTC}}_{\mathrm{fail},t}]
\leq0.03,\quad \forall t,
\tag{C2-I}\label{eq:c2}\\
&\frac{1}{T}\sum_t\mathbb E[\mathrm{CBR}_{t}]
\leq0.65,
\tag{C3-L}\label{eq:c3}\\
&\frac{1}{NT}\sum_{n,t}\mathbb E\!
\left[\frac{R^{\Uu}_{\min}-R^{\Uu}_{n,t}}{R_0}\right]
\leq0,
\tag{C4-L}\label{eq:c4}\\
&\sum_{r\in\Rsl}y_{n,r,t}=1,\quad
\forall n\in\Nset,\ t,\nonumber\\[-2pt]
&y_{n,r,t}\in\{0,1\},\quad
\forall n\in\Nset,\ r\in\Rsl,\ t,
\tag{C5-H}\label{eq:c5}\\
&P^{\SL}_{n,t}\in\Psl,\quad\forall n\in\Nset,\ t,
\tag{C6-H}\label{eq:c6}\\
&y_{n,r,t}+y_{m,r,t}\leq1,\quad
\forall r\in\Rsl,\ t,\nonumber\\[-2pt]
&\quad (n,m)\in\Ecal^{\mathrm{prot}}_t,
\tag{C7-C}\label{eq:c7}\\
&\mathcal F_{n,t}\ne\varnothing\Longrightarrow\nonumber\\[-2pt]
&\quad\Pi_n(\tilde a_{n,t};\Gcal^{\mathrm{ego}}_{n,t},
\widetilde{\mathcal P}_{n,t})
\in\mathcal F_{n,t},\nonumber\\[-2pt]
&\quad\forall n\in\Nset,\ t.
\tag{C8-C}\label{eq:c8}
\end{align}
\endgroup
Here \(y_{n,r,t}=\ind\{\rho^{\SL}_{n,t}=r\}\), \(\tilde a_{n,t}\) is the
actor proposal, and \(\Pi_n\) is the local filter.  Expectations cover policy
sampling, mobility initialization, shadowing, fading, and exogenous Uu grants.
The 3\% low-TTC budget, compared with 10\% aggregate packet failure, assigns a
stricter communication target to route-conflicting pairs.  During repair-free
training, \(\Pi_n\) is the identity; the nontrivial filter is introduced only
for filtered evaluation and deployment.
The protected-edge set
\(\Ecal^{\mathrm{prot}}_t\) contains mutually observed one-hop low-TTC or
local-interference pairs to which resource separation is applied, and
\(\mathcal F_{n,t}\) is vehicle \(n\)'s locally feasible decoded-action set.
The labels record how each condition is enforced.  C1, C3, and C4 are
long-run expected budgets.  C2 is an expected per-epoch requirement that
prevents low-TTC failures in one part of an episode from being averaged away
by benign epochs.  It remains statistical rather than a sample-path guarantee.
The action decoder enforces the structural conditions C5--C6 at every epoch,
independently of the learned policy or filter.  C7--C8 describe
local execution rules that depend on fresh
one-hop information and a nonempty \(\mathcal F_{n,t}\); a valid fallback is
used otherwise.  This hierarchy keeps statistical regulation, action
validity, and deployment repair conceptually separate.  Physical-risk
variables determine communication priority within this radio-control problem.

\section{PA-STGAT Constrained Reinforcement Learning}
\label{sec:method}

PA-STGAT uses centralized training and decentralized execution (CTDE) to link
local radio decisions to delayed, network-level outcomes.  Its agentic loop is
specific to this task.  Each vehicle perceives a one-hop, multi-relation ego
graph, retains four recent snapshots, and acts with its own copy of a shared
policy.  During training, centralized critics evaluate the joint action and
four future costs; measured failures update the AL prices.  Credit assigns
delivery outcomes to transmitters, while a PDR-priced graph surrogate puts a
direct coordination gradient on categorical resource choices.  The next
epoch supplies a new graph and another local decision.  At execution, no
global graph, centralized critic, dual update, or central PC5 scheduler is
needed; the local filter uses only one-hop control messages.

\subsection{Dynamic Multi-Relation Ego Graph}
\label{subsec:graph}

A neighbor can play different roles: it may be an intended receiver, a source
of interference, or a route-conflicting vehicle that deserves elevated radio
priority.  One adjacency matrix would erase these distinctions.  We retain
them as three relations on a common epoch-wise graph.  Centralized training
maintains
\begin{equation}
\Gcal_t=(\Nset,X_t,\Ecal_t^{\mathrm{comm}},
\Ecal_t^{\mathrm{int}},\Ecal_t^{\mathrm{risk}}),
\label{eq:global_graph}
\end{equation}
where \(X_t=[\bm x_{1,t},\ldots,\bm x_{N,t}]^{\mathsf T}\) stacks node
features.  The communication relation connects observable PC5 neighbors, the
interference relation connects transmitters whose current assignments can
couple, and the risk relation connects route-conflicting pairs within the
\(\tau_{\mathrm{ref}}\) horizon.  Risk edges carry the priority
\(w_{n,m,t}\) from~\eqref{eq:risk_weight}.  Every 10-ms action and motion
update rebuilds all three edge sets.  The value \(w_{n,m,t}\) is an input
feature; the attention coefficients in~\eqref{eq:gat} learn how strongly it
affects the node embedding.

The edge features describe why a relation exists.  Communication edges encode
link gain and distance; interference edges encode received power and recent
resource overlap; risk edges encode TTC, priority, and the route gate.  Let
\(\widehat g^{\SL}_{i,j,t}\) be the current large-scale PC5 gain estimate,
let \(d_{\mathrm{comm}}\) be the communication range, and let
\(\bar\rho^{\SL}_{i,t}\) and \(\bar P^{\SL}_{i,t}\) denote the most
recently executed PC5 resource and power observed at epoch \(t\).  Then
\begin{align}
\bm f^{\mathrm{comm}}_{i,j,t}
&=\left[\log_{10}(\widehat g^{\SL}_{i,j,t}+\varepsilon_0),\
\frac{d_{i,j,t}}{d_{\mathrm{comm}}}\right]^{\mathsf T},\nonumber\\
\bm f^{\mathrm{int}}_{i,j,t}
&=\left[\log_{10}(\bar P^{\SL}_{i,t}
\widehat g^{\SL}_{i,j,t}+\varepsilon_0),\
\ind\{\bar\rho^{\SL}_{i,t}=\bar\rho^{\SL}_{j,t}\}\right]^{\mathsf T},
\nonumber\\
\bm f^{\mathrm{risk}}_{i,j,t}
&=\left[\frac{\min(\tau^{\mathrm{rad}}_{i,j,t},\tau_{\mathrm{ref}})}
{\tau_{\mathrm{ref}}},\
\frac{w_{i,j,t}}{1+\alpha_r w_{\max}},\
\chi_{i,j,t}\right]^{\mathsf T}.
\label{eq:edge_features}
\end{align}
Interference edges are instantiated only for same-resource pairs within the
interference range.  The communication, interference, and risk vectors have
dimensions two, two, and three, respectively.

At deployment, each vehicle reconstructs the graph from information it has
actually received.  At the beginning of epoch \(t\), vehicle \(n\) constructs
\begin{equation}
\Gcal^{\mathrm{ego}}_{n,t}
=\Gcal_t[\{n\}\cup\Nset^{\mathrm{obs}}_{n,t}],
\label{eq:ego_graph}
\end{equation}
from its state and the latest metadata received over a logical low-rate
state-beacon channel.  Each beacon carries timestamped position and velocity,
the latest executed PC5 resource and power, the Uu grant state, one-hop
large-scale gain estimates, and the TTC-derived descriptors in
\eqref{eq:edge_features}.  These measurements are the inputs to local graph
inference; vehicles preload the shared ST-GAT parameters and exchange no
learned embeddings.  Missing records remove the associated nodes and edges,
and age and validity masks expose stale information to the encoder.

This information boundary also explains the role of CTDE.  The actor observes
beacon neighbors within 120 m.  The training graph extends interference edges
to 180 m.  Hidden same-resource transmitters in that
outer region can affect the centralized critic's target even though they are
absent from the deployed actor's ego graph.

Beacon reception occurs before actor inference and uses a control process
outside the 14-resource payload action pool.  The current payload choice
therefore cannot alter the graph used to produce that same choice.  A
preconfigured NR sidelink resource subpool provides one possible realization
~\cite{3gpp38214,3gpp38321}.  In the simulator, missing or stale beacons are
exogenous observation impairments.  Beacon overhead, beacon--payload spectral
leakage, and delivery guarantees remain outside the model.

Within each relation, attention learns which observed neighbor is most
informative for the current action.  Separate parameters keep communication,
interference, and risk from sharing one aggregation rule.  We use the
rectified linear unit (ReLU) and its leaky variant (LeakyReLU).  For relation
\(\ell\in\{\mathrm{comm},\mathrm{int},\mathrm{risk}\}\), a GAT head computes
\begin{align}
\xi^{\ell}_{i,j,t}
&=\mathrm{LeakyReLU}\!\left(
\left(\bm A_{\ell}^{\mathrm{att}}\right)^{\mathsf T}
[W_\ell\bm x_{i,t}\Vert W_\ell\bm x_{j,t}
\Vert\bm f^\ell_{i,j,t}]\right),\\
\mathrm{att}^\ell_{i,j,t}
&=\frac{\exp(\xi^\ell_{i,j,t})}
{\sum_{k\in\mathcal N^\ell_i}\exp(\xi^\ell_{i,k,t})},\\
\bm h_{i,t}
&=\mathrm{ReLU}\!\left(W_0\bm x_{i,t}
+\sum_\ell\sum_{j\in\mathcal N^\ell_i}
\mathrm{att}^\ell_{i,j,t}W_\ell\bm x_{j,t}\right).
\label{eq:gat}
\end{align}
Here \(\mathcal N_i^\ell=\{j:(i,j)\in\Ecal_t^\ell\}\) is node \(i\)'s
relation-\(\ell\) neighborhood; \(W_\ell\) and \(W_0\) are learnable
feature projections, \(\bm A_\ell^{\mathrm{att}}\) is the corresponding
attention vector, and \(\Vert\) denotes concatenation.
The relation-specific parameters distinguish physical-risk edges from
communication and interference edges.

\begin{figure}[t]
    \centering
    \includegraphics[width=0.85\columnwidth]{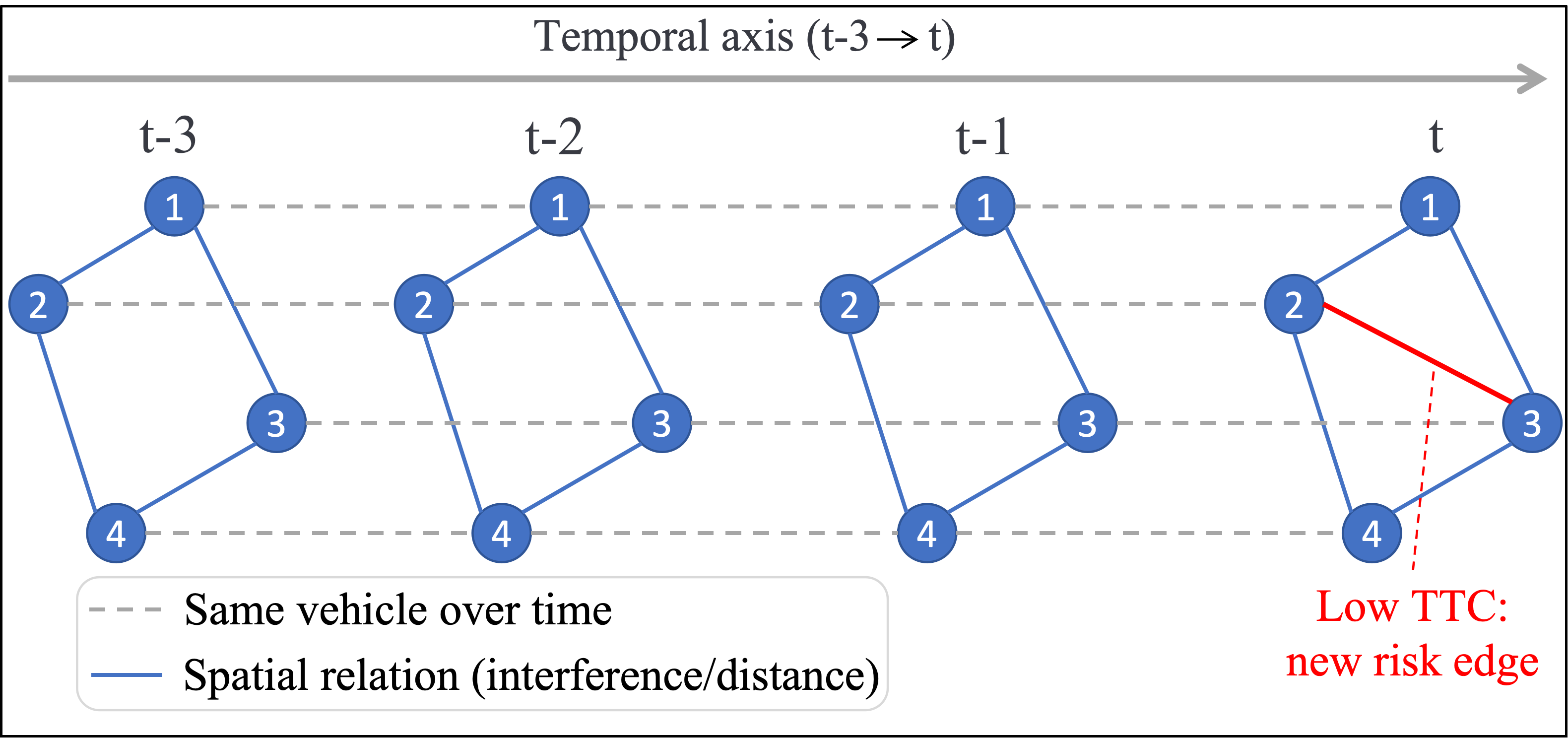}
    \caption{Evolution of spatial communication/interference relations and
    TTC-derived physical-risk relations over the \(K=4\) history
    \((t-3,\ldots,t)\). Relation-specific GAT layers and a GRU form the actor
    embedding; the current relation sets are also reused by the graph-conflict
    surrogate in Section~\ref{subsec:al}.}
    \label{fig:stgat}
    \vspace{-12pt}
\end{figure}

Rebuilding the graph follows topology changes, but a single snapshot remains
sensitive to fading and message age.  A shared gated recurrent unit (GRU)
therefore links \(K=4\) consecutive spatial embeddings,
\begin{equation}
\bm z^{\mathrm{temp}}_{i,t}
=\mathrm{GRU}(\bm h_{i,t},\bm z^{\mathrm{temp}}_{i,t-1}),
\qquad
\bm z_{n,t}=\bm z^{\mathrm{temp}}_{n,t},
\label{eq:gru}
\end{equation}
and \(\bm z_{n,t}\) enters vehicle \(n\)'s actor.  The four snapshots span
30 ms.  Each snapshot uses its own reconstructed edges, so the GRU summarizes
recent evolution without freezing topology across the window.
This is short-term state memory, not an online planner or a mechanism for
updating policy parameters after deployment.
Fig.~\ref{fig:stgat} depicts this sequence.  The graph then has two downstream
roles: its temporal ego embedding drives the shared actor, and its current
communication and risk edges define the PDR-priced conflict signal.

\subsection{Shared Actor and Centralized Critics}
\label{subsec:ctde}

Homogeneous vehicles solve the same local mapping from an ego graph to a radio
action, so they share one actor parameter set,
\begin{equation}
\pi_\theta(a_{n,t}\mid
\Gcal^{\mathrm{ego}}_{n,t-K+1:t})
=\pi_\theta(a_{n,t}\mid\bm z_{n,t}).
\label{eq:shared_actor}
\end{equation}
During training, the \(N\) ego graphs form a batch and one optimizer updates
\(\theta\) from the mean per-vehicle loss.  Deployment copies the resulting
actor and ST-GAT weights to every vehicle; each copy processes its own local
graph.  The training-time optimizer produces one parameter set; it does not
average separately trained vehicle models.  Deployment needs neither a model
server nor a centralized action selector.  This is parameter sharing within
CTDE, with no federated aggregation.

A local actor cannot observe every interferer or reconstruct a network-level
failure by itself.  The centralized critics provide that training context by
observing the global graph and joint decoded action.
Let \(\bm a_t=(a_{1,t},\ldots,a_{N,t})\) denote that joint action.
Their shared heads return one utility estimate and four cost estimates for
each vehicle,
\(Q^u_{\psi_u,n}(s_t,\bm a_t)\in\mathbb R\) and
\(\bm Q^c_{\psi_c,n}(s_t,\bm a_t)\in\mathbb R^4\); stacking over vehicles
gives outputs in \(\mathbb R^N\) and \(\mathbb R^{N\times4}\), respectively.
The global context lets these per-vehicle heads evaluate a joint decision, and
their mean utility recovers \(u_t\) in~\eqref{eq:utility}.  The shared
multi-agent deep deterministic policy gradient (MADDPG) actor outputs a
resource softmax and sigmoid-normalized power.  Execution takes the resource
argmax and quantizes power to the nearest level in \(\Psl\); training retains
the continuous outputs for differentiation.  During training, replay samples
and target networks update the utility and cost critics from measured utility,
transmitter-level costs, and subsequent joint actions.  The updates average
over vehicles and account for episode termination.  This CTDE structure
is consistent with recent MADDPG coordination studies~\cite{zhu2026maddpg}.
The remaining question is how a network-level PDR failure becomes a meaningful
target for each vehicle's cost head.

\subsection{Metric-Preserving Constraint Credit}
\label{subsec:credit}

Giving every vehicle the same packet-failure scalar would make their cost
targets indistinguishable.  We assign each transmitter the failures observed
on its outgoing intended links and normalize the assignments so that their
vehicle average preserves the original network metric:
\begin{align}
c^{\mathrm{PDR}}_{n,t}
&=\frac{N\sum_m M_{n,m,t}(1-S_{n,m,t})}
{\max(\sum_{i,j}M_{i,j,t},1)},\nonumber\\
c^{\mathrm{TTC}}_{n,t}
&=\frac{N\sum_m Z_{n,m,t}(1-S_{n,m,t})}
{\max(\sum_{i,j}Z_{i,j,t},1)}.
\label{eq:credits}
\end{align}
The complete per-vehicle vector is
\begin{equation}
\bm c_{n,t}=
\begin{bmatrix}
c^{\mathrm{PDR}}_{n,t}\\
c^{\mathrm{TTC}}_{n,t}\\
\mathrm{CBR}_t\\
(R^{\Uu}_{\min}-R^{\Uu}_{n,t})/R_0
\end{bmatrix},
\qquad
\bm c_t=[\bm c_{1,t},\ldots,\bm c_{N,t}]^{\mathsf T}.
\label{eq:cost_vector}
\end{equation}
The first two entries assign link outcomes to a transmitter.  CBR is copied to
all vehicles because occupancy is a shared network-load measurement, and the
Uu residual remains vehicle-specific.  Their average satisfies
\begin{equation}
\frac{1}{N}\sum_n\bm c_{n,t}
=
\begin{bmatrix}
1-\mathrm{PDR}_t\\
p^{\mathrm{TTC}}_{\mathrm{fail},t}\\
\mathrm{CBR}_t\\
\frac{1}{N}\sum_n(R^{\Uu}_{\min}-R^{\Uu}_{n,t})/R_0
\end{bmatrix}.
\label{eq:credit_identity}
\end{equation}
Equation~\eqref{eq:credit_identity} guarantees accounting consistency between
the critic targets and the CMDP budgets.  Interaction remains a joint-action
effect: the centralized critic observes all actions, and
\(\mathcal L_{\mathrm{gconf}}\) in~\eqref{eq:gconf} couples the resource
probabilities of potential interferers.  Credit therefore localizes measured
outcomes, and the graph term supplies the coordination signal.

\subsection{Augmented Lagrangian Actor Update}
\label{subsec:al}

%%%%%%%%%%%%%%%%%%%%%%%%%%%%%%%%%%%%
%%%%%%%%%%%%%%%%%%%%%%%%%%%%%%%%%%%%
%%%%%%%%%%%%%%%%%%%%%%%%%%%%%%%%%%%%

The cost critics predict how the current joint action affects future budgets.
AL turns each predicted excess into an adaptive price in the actor update.
Let \(\bm\epsilon=[0.10,0.03,0.65,0]^{\mathsf T}\) collect the budgets in the order of \(\bm c_{n,t}\), and define
\begin{equation}
\bm\lambda
=[\lambda_{\mathrm{PDR}},\lambda_{\mathrm{TTC}},
\lambda_{\mathrm{CBR}},\lambda_{\Uu}]^{\mathsf T},
\label{eq:al_components}
\end{equation}
where \(\bm\lambda\) contains nonnegative dual prices.  The cost critic uses
discounted targets for the long-run constraints
\(\mathcal J_L=\{\mathrm{PDR},\mathrm{CBR},\Uu\}\) and a one-step target for
the instantaneous TTC component.  The corresponding actor residual is
\begin{equation}
\varphi_{j,n}(s,\bm a)=
\begin{cases}
Q^c_{\psi_c,n,j}(s,\bm a)-\epsilon_j,
&j\in\mathcal J_L,\\
[Q^c_{\psi_c,n,\mathrm{TTC}}(s,\bm a)-\epsilon_{\mathrm{TTC}}]_+,
&j=\mathrm{TTC}.
\end{cases}
\label{eq:residual}
\end{equation}
Thus, a short TTC violation is priced at its current epoch, while the other
three heads estimate long-run cost.  

The proposed PA-STGAT MADDPG actor minimizes
\begin{align}
\mathcal L(\theta)
=\mathbb E_{
% s_t\sim\mathcal D
\pi(\theta)
}
&\Bigg[
\frac{1}{N}\sum_{n=1}^{N}\Big(
-Q^u_{\psi_u,n}(s_t,\bm a_t^\pi)
+\bm\lambda^{\mathsf T}
 \bm\varphi_{\psi_c,n}(s_t,\bm a_t^\pi)
\nonumber\\
&+\frac{\zeta_{\mathrm{TTC}}}{2}
 {\varphi_{\mathrm{TTC},n}
 (s_t,\bm a_t^\pi)}^2
\Big)
+\Phi^{\mathrm{AL}}_{\mathrm{gconf},t}
\Bigg].
\label{eq:actor_loss}
\end{align}

Here, $\mathcal D$ is the replay distribution, and $\bm a_t^\pi$ is the joint action recomputed by the current actors at the sampled state $s_t$. 
The factor $1/N$ averages the vehicle-specific critic outputs; the graph-conflict term is shared across vehicles and is added once per sampled graph.
The first term promotes Uu utility; the next two penalize
predicted cost excess according to its current dual price.  The quadratic term
is reserved for the instantaneous TTC constraint; C1, C3, and C4 contain no
quadratic AL penalty.  The
environment continues to report \(u_t\) from~\eqref{eq:utility}, and the
constraint terms enter through this learning objective.  In particular,
\(\lambda_{\mathrm{TTC}}\) and \(\zeta_{\mathrm{TTC}}\) price the predicted
one-step low-TTC residual.

This AL path is still indirect for a categorical resource softmax: the actor
must infer from the critic how two vehicles' probabilities create a collision.
We expose that relation with a differentiable same-resource surrogate.
Let \(\pi_{\theta,n}(r)\) denote vehicle \(n\)'s marginal probability of
selecting PC5 resource \(r\).  For relation type
\(\upsilon\in\{\mathrm{risk},\mathrm{comm}\}\), define
\begin{align}
\mathcal C^\upsilon_t
&=\frac{\sum_{(n,m)\in\Ecal^\upsilon_t}\omega^\upsilon_{n,m,t}
\sum_{r\in\Rsl}\pi_{\theta,n}(r)\pi_{\theta,m}(r)}
{\max(\sum_{(n,m)\in\Ecal^\upsilon_t}\omega^\upsilon_{n,m,t},\varepsilon_0)},
\nonumber\\
\omega^{\mathrm{risk}}_{n,m,t}
&=1+2\left(1-\frac{\min(\tau^{\mathrm{rad}}_{n,m,t},
\tau_{\mathrm{ref}})}{\tau_{\mathrm{ref}}}\right),\quad
\omega^{\mathrm{comm}}_{n,m,t}=1,\nonumber\\
\mathcal L_{\mathrm{gconf}}
&=\mathcal C^{\mathrm{risk}}_t+0.3\mathcal C^{\mathrm{comm}}_t.
\label{eq:gconf}
\end{align}
Physical risk now enters a second path.  The encoder treats
\(w_{n,m,t}\) from~\eqref{eq:risk_weight} as a learned-attention feature;
the surrogate uses \(\omega^{\mathrm{risk}}_{n,m,t}\) as a fixed linear weight
on same-resource overlap.  The conflict is priced by the measured-PDR dual,
\begin{equation}
\begin{aligned}
\Phi^{\mathrm{AL}}_{\mathrm{gconf}}
&=\beta_g\!\left[
\lambda_{\mathrm{PDR}}\mathcal L_{\mathrm{gconf}}
+\frac{\rho_g}{2}\mathcal L_{\mathrm{gconf}}^2\right],\\[-2pt]
&\hspace{22mm}\beta_g=1,\qquad \rho_g=0.1.
\end{aligned}
\label{eq:gconf_al}
\end{equation}
Equation~\eqref{eq:gconf_al} couples the three signals: current edges select the
pairs, the explicit risk weight sets relative priority, and the adaptive PDR
price sets gradient strength.  PDR prices the surrogate because same-resource
overlap approximates aggregate delivery conflict.  The fixed \(\rho_g\) is a
curvature term for this instantaneous conflict surrogate, not a quadratic
penalty on the cumulative PDR budget.  The low-TTC budget retains its own
\(\lambda_{\mathrm{TTC}}\) and \(\zeta_{\mathrm{TTC}}\) in
\eqref{eq:actor_loss}, combining its critic signal with risk-prioritized
categorical coordination.

The surrogate operates on marginal resource probabilities, so it is a smooth
upper-level approximation to conflicts after argmax decoding.  Its purpose is
to expose a usable gradient; measured episode costs determine the price.  The
dual vector is updated once per episode from the exponentially smoothed cost:
\begin{equation}
\bm\lambda\leftarrow
\mathrm{clip}\!\left(
\bm\lambda+\eta_\lambda(\widehat{\bm J}^{c}-\bm\epsilon),
\bm0,\bm\lambda_{\max}\right).
\label{eq:dual}
\end{equation}
For C1, C3, and C4, \(\widehat J^{c}_j\) is the exponential moving average
of the episode mean \(T^{-1}\sum_tN^{-1}\sum_nc^j_{n,t}\).  For C2, it is
the exponential moving average of
\(\max_tN^{-1}\sum_nc^{\mathrm{TTC}}_{n,t}\), a conservative empirical
statistic for the per-epoch requirement.  Here \(\eta_\lambda\) is the dual
step size, and
\(\mathrm{clip}(\cdot,\bm0,\bm\lambda_{\max})\) applies componentwise
nonnegative bounds.
This update occurs during offline training.  The deployed actor receives
neither measured network-wide costs nor a running dual variable; constraint
feedback has already shaped its fixed policy parameters.
Only the instantaneous TTC coefficient is increased while its smoothed
residual remains positive:
\begin{equation}
\zeta_{\mathrm{TTC}}\leftarrow
\begin{cases}
\min\{\zeta_{\max},1.01\zeta_{\mathrm{TTC}}\},
&\widehat J^c_{\mathrm{TTC}}-\epsilon_{\mathrm{TTC}}>10^{-4},\\
\zeta_{\mathrm{TTC}},&\text{otherwise}.
\end{cases}
\label{eq:zeta_update}
\end{equation}

\subsection{Fixed-Penalty Reward-Shaping Baseline}
\label{subsec:noal}

To test adaptive pricing against coefficient tuning, we use a matched No-AL
baseline with the same observations, ST-GAT actor--critic, replay budget, and
metric-preserving costs.  The comparison changes how constraints enter
learning.  PA-STGAT keeps the replay reward equal to Uu utility and gives each
budget its own adaptive AL price; No-AL replaces those updates with fixed
coefficients and trains the utility critic on the shaped reward
\begin{align}
\mathsf{Pen}^{\mathrm{fixed}}_{n,t}
&=\bar{\bm\nu}^{\mathsf T}
\pos{\bm c_{n,t}-\bm\epsilon}
+\bar\nu_g\ell^{\mathrm{gconf}}_{n,t},\\
u^{\mathrm{shp}}_{n,t}
&=u_{n,t}-\mathsf{Pen}^{\mathrm{fixed}}_{n,t}.
\label{eq:noal}
\end{align}
A local term \(\ell^{\mathrm{gconf}}_{n,t}\) assigns the graph-conflict
penalty to vehicle \(n\) from its incident
risk and communication edges using the same resource-probability overlap
as~\eqref{eq:gconf}.  The fixed vectors \(\bar{\bm\nu}\) and
\(\bar\nu_g\) price measured cost excess and local graph conflict,
respectively.
Four coefficient settings were screened for 250 episodes on one development
seed without the projection auxiliary.  The selected values,
\(\bar{\bm\nu}=[5,2,1,1]^{\mathsf T}\) and \(\bar\nu_g=5\), were then fixed
across all evaluation seeds.
The matched architecture makes adaptive pricing the sole learning difference.
Both policies execute raw actions during training, so their replay buffers
record the consequences of the learned proposals.

\subsection{Repair-Free Training and Local Filtering}
\label{subsec:filter}

Applying a repair before storing experience would make a poor proposal appear
successful.  Training therefore executes the decoded raw action and records
its actual consequence.  Deployment adds the local repair map.  Specifically,
\(\widetilde{\mathcal P}_{n,t}\) contains the decoded resource--power proposals
and resource-preference vectors received from mutually observed neighbors after
actor inference.  For both AL and fixed-penalty policies,
\begin{equation}
a^{\mathrm{train}}_{n,t}=\tilde a_{n,t},\qquad
a^{\mathrm{dep}}_{n,t}
=\Pi_n(\tilde a_{n,t};\Gcal^{\mathrm{ego}}_{n,t},
\widetilde{\mathcal P}_{n,t}).
\label{eq:split}
\end{equation}
Mandatory decoding applies in both phases; \(\Pi_n\) is activated for filtered
evaluation and deployment.

The filter resolves local conflicts in two stages.  First, a deterministic
order lets mutually observed protected neighbors announce assignments in
sequence.  Protection covers low-TTC intended pairs and local interference
pairs.  Let \(\mathcal B_{n,t}\) contain the resources already announced by
higher-ranked protected neighbors.  Second, the filter evaluates the remaining
resource--power candidates by their predicted link success.  Let
\(\widehat q_{n,t}(r,p)\) be the minimum predicted success probability over
vehicle \(n\)'s observed intended receivers.  The local feasible set
\(\mathcal F_{n,t}\) contains \((r,p)\in\Rsl\times\Psl\) satisfying
\(r\notin\mathcal B_{n,t}\) and \(\widehat q_{n,t}(r,p)\geq p_0\).
For each of the up to
\(|\Rsl||\Psl|=42\) nominal candidates, the filter estimates an independent-Rayleigh
success surrogate using locally known PC5 interferers, exogenous Uu
interference, and
\begin{equation}
\Gamma_{\mathrm{eff}}=
\max\!\left\{\Gamma^{\SL},
2^{L_{\mathrm{pkt}}/(B^{\SL}D_{\max})}-1\right\}
10^{\Delta_\Gamma/10},
\label{eq:filter_threshold}
\end{equation}
where \(\Delta_\Gamma=3\) dB is the SINR margin.  Half-duplex candidates
receive zero reliability.
For the nominal \(2400\)-bit packet, \(180\)-kHz bandwidth, and \(10\)-ms
budget, the delay-derived threshold is \(1.82\) dB, below the \(3\)-dB
decoder threshold.  The latter therefore dominates before the margin,
giving a nominal effective threshold of \(6\) dB.
Here \(p_0=0.90\).  A nonempty \(\mathcal F_{n,t}\) is ranked by power,
predicted reliability, and actor preference.  If the reliability gate rejects
all powers, the filter uses the highest available power and the best residual
resource score.  If neighbor protection exhausts the resource set, it reopens
the decoded candidates.  These fallbacks always produce a valid action, with
conditional reliability and separation.

Execution uses ego-graph records and one-hop control messages.  The simulator
implements proposal and assignment exchange as a reliable deterministic
routine completed within the epoch; Section~\ref{subsec:stress} perturbs the
preceding state beacons.  No center arbitrates PC5 assignments, although
filtered execution still depends on neighbor signaling.  The resulting
guarantees are deliberately narrow:
decoding enforces C5--C6, and C7--C8 depend on fresh information and a nonempty
feasible set.  The local reliability score is an uncalibrated predictor, and
the filter does not evaluate the trajectory-level CBR or Uu budgets.  C1--C4
therefore retain statistical meanings: C2 is per-epoch in expectation, while
C1, C3, and C4 are long-run averages.  Algorithm~\ref{alg:training}
summarizes the learning and deployment paths.

\begin{algorithm}[t]
\caption{PA-STGAT Optimization and Decentralized Execution}
\label{alg:training}
\begin{algorithmic}[1]
\STATE Initialize actor, centralized critics, targets, replay,
\(\bm\lambda=\bm0\), \(\zeta_{\mathrm{TTC}}=\zeta_0\), and
\(\rho_g=0.1\).
\FOR{each training episode}
  \FOR{each 10-ms epoch}
    \STATE Rebuild global relations and each masked one-hop ego graph.
    \STATE Encode \(K\) snapshots with relation GATs and GRU.\@
    \STATE Sample/decode raw actions; execute without reliability repair.
    \STATE Store utility, metric-preserving costs, graphs, and actions.
    \STATE Update critics and actor using~\eqref{eq:actor_loss}.
  \ENDFOR
  \STATE Update \(\bm\lambda\) and \(\zeta_{\mathrm{TTC}}\) from measured costs
  using~\eqref{eq:dual}--\eqref{eq:zeta_update}.
\ENDFOR
\STATE Deploy identical actor weights to every vehicle; discard critics and duals.
\FOR{each deployment epoch, independently at each vehicle}
  \STATE Receive one-hop beacons and rebuild the recent ego graphs.
  \STATE Infer and decode a local PC5 resource--power proposal.
  \STATE Exchange one-hop proposals and apply \(\Pi_n\) without a central controller.
\ENDFOR
\end{algorithmic}
\end{algorithm}

\subsection{Feasibility and Near-Optimality Analysis}
We now analyze the AL training process for the statistical constraints C1--C4 in Problem~\eqref{eq:problem}. To keep the analysis compact, we use an abstract notation that matches the formulation in Appendix~\ref{app:sec:analysis}. Specifically, \(r_t\) denotes the reward, or utility in the main text. The vector \(c_t\) collects the residuals of the long-run constraints C1, C3, and C4, while \(g_t\) denotes the residual of the per-epoch C2 constraint. For a policy \(\pi\), \(J(\pi)\) is the expected discounted utility, \(C(\pi)\) is the average long-run constraint residual, and \(H(\pi)\) and \(Q(\pi)\) measure the positive C2 violation and its quadratic penalty, respectively. The analysis shows that, under the assumptions below, the AL iterations approach feasibility while the achieved utility remains within the policy-solver error of the constrained optimum.

\noindent\textbf{Assumption 1 (bounded and closed performance).}
The functions $r_t,c_t,g_t$ are bounded, and the attainable performance set $\mathcal V=\{(J(\pi),C(\pi),H(\pi),Q(\pi)):\pi\in\Pi\}$
is compact, and the class $\Pi$ is closed. 
% under finite episode-level mixtures.
% one policy is selected at the start of an episode and executed throughout
% that episode. 

% \medskip
\noindent\textbf{Assumption 2 (instantaneously feasible optimum).}
Problem~\eqref{app:eq:problem} admits an optimizer $\pi^*$ satisfying
$H(\pi^*)=0$. Equivalently, $g_t\leq0$ almost surely under $\pi^*$ at
every $t<T$. This is the finite-horizon counterpart of the pointwise
optimal-policy condition used in the augmented-Lagrangian analysis of
Li et al.~\cite[Proposition~3]{li2021alm}.

% Since all $\gamma^t$ are positive, this condition gives $Q(\pi^*)=0$ and preserves the original constrained reward benchmark when instantaneous residuals are clipped.

% \medskip
\noindent\textbf{Assumption 3 ($\delta$-capable policy solver).}
We assume that there exists some constant $\delta\geq0$ such that $\forall \lambda_k, \rho_k$, the policy-searching subroutine returns some
$\pi_k\in\Pi$ such that
\begin{equation}
 R_{\rho_k}(\pi_k,\lambda_k)
 \geq\max_{\pi\in\Pi}R_{\rho_k}(\pi,\lambda_k)-\delta.
\label{app:eq:capability_define}
\end{equation}
Here, $R_\rho$ represents the training objective based on the augmented Lagrangian method, as the one defined in \eqref{eq:actor_loss}. 
The same $\delta$ applies to every iteration $k$. In PA-STGAT,
$\pi_k=\pi_{\theta_k}$, and this condition measures the solver's accuracy
in augmented-objective value. Actor approximation, critic estimation, and
the effect of the graph-conflict training term are included in this
accuracy requirement for~\eqref{app:eq:surrogate}.

\begin{theorem}[Uniform-accuracy iteration guarantee]
\label{app:thm:main_define}
Under Assumptions~1--3 and
\eqref{app:eq:eta}--\eqref{app:eq:penalty}, the uniform episode mixture $F(\bar\pi_K) \triangleq \frac{1}{K} \sum_{k=0}^{K-1}F(\pi_k)$, where $F \in \{J,C, H,Q \}$, satisfies, for every $K $, 
\begin{equation}
 J(\bar\pi_K)\geq J^*-\delta-\varepsilon_K,
\label{app:eq:finite-reward_define}
\end{equation}
where $\varepsilon_K
 =\frac{dM_{K-1}^2}{2K\eta_{K-1}}
   +\frac{B^2}{2K}\sum_{k=0}^{K-1}\eta_k.$
For the fixed horizon $T$, as $K\to\infty$, $ \varepsilon_K \to 0$. Moreover, 
\begin{gather}
 \begin{gathered}
 [C(\bar\pi_K)]_+\longrightarrow0,\\
 H(\bar\pi_K)\longrightarrow0,\qquad Q(\bar\pi_K)\longrightarrow0,
 \end{gathered}
 \label{app:eq:feasibility_define}\\
 [G_t(\bar\pi_K)]_+\longrightarrow0,
 \qquad t=0,\ldots,T-1,
 \label{app:eq:instantaneous_define}\\
 J^*-\delta\leq\liminf_{K\to\infty}J(\bar\pi_K)
 \leq\limsup_{K\to\infty}J(\bar\pi_K)\leq J^*.
 \label{app:eq:reward-limit_define}
\end{gather}
Thus a $\delta$-capable solver achieves asymptotic feasibility with a
reward gap of at most $\delta$. In particular, $\delta=0$ gives
$J(\bar\pi_K)\to J^*$.
\end{theorem}
\begin{IEEEproof}
    The proof is given in Appendix~\ref{app:sec:analysis}.
\end{IEEEproof}

\section{Experimental Evaluation}
\label{sec:experiments}

We compare PA-STGAT with learned baselines and then examine its architecture, constraint pricing, scalability, robustness, and local filtering. The C2 budget is \(0.03\). C2 uses the instantaneous quadratic penalty, while C1, C3, and C4 remain cumulative constraints without quadratic penalties.

\subsection{Setting and Metrics}
\label{subsec:setup}

The nominal environment has eight vehicles, 14 PC5 resources, and three exogenous Uu grants. Each vehicle selects one resource and power for a V2V packet every 10 ms. Each episode contains 100 epochs. Vehicle speeds are 8--16 m/s. The communication and interference ranges are 120 and 180 m. PC5 and Uu bandwidths are 180 and 360 kHz, respectively. Packets contain 2400 bits with a 10-ms delay budget and a 3-dB decoding threshold. The channel includes Rayleigh fading, 3-dB shadowing, path-loss exponents 2.2/2.5, and leakage coefficients \(\kappa_{\Uu\rightarrow\SL}=0.02\) and \(\kappa_{\SL\rightarrow\Uu}=0.05\). The actor uses two attention heads, a 48-dimensional hidden state, and four graph snapshots. Training executes decoded raw actions. The local filter is applied only during filtered evaluation.

Five independent training seeds \(\{7,17,29,43,61\}\) are used. Reported uncertainties are 95\% Student-\(t\) confidence half-widths across seeds. PDR, channel busy ratio (CBR), and low-TTC failure are reported as episode means. For C2, low-TTC failure is first averaged across held-out episodes at each epoch. The largest epoch-wise mean is reported as C2 max. 
A seed is feasible when filtered PDR is at least \(0.90\), CBR is at most \(0.65\), the Uu residual is nonpositive, and C2 max is below \(0.03\). 
Feasibility is an empirical classification on the specified held-out set. The maximum over epoch-wise sample means makes the C2 estimate sensitive to evaluation sampling near the threshold; feasible-seed counts do not certify constraint satisfaction on unseen trajectories. Resource repair denotes the fraction of vehicle resources modified by the filter within an epoch, rather than resource switching between consecutive epochs. It is zero by construction during raw execution.
The Uu residual is \(0.20-U/100\), where \(U=\sum_tu_t\). At eight vehicles, \(\mathrm{CBR}_t\leq8/14<0.65\) by construction. CBR regulation is therefore better exposed by the vehicle-count experiments. The overlap and leakage model, together with the omitted beacon overhead, makes this a system-level evaluation rather than a protocol-calibrated performance estimate.

\begin{table}[t]
\centering
\caption{Common mixed-constraint training settings.}
\label{tab:learning_parameters}
\footnotesize
\begin{tabular}{@{}p{0.51\columnwidth}p{0.41\columnwidth}@{}}
\toprule
Parameter & Value\\
\midrule
Discount / Polyak factor & \(0.95/0.01\)\\
Actor or Q-network / critic learning rate & \(2{\times}10^{-4}/3{\times}10^{-4}\)\\
Batch / replay capacity & \(16/10{,}000\)\\
Learning start / update interval & \(100/10\) epochs\\
Dual step / cap & \(0.05/25\)\\
C2 quadratic initial / growth / cap & \(0.1/1.01/10\)\\
Graph-conflict curvature & \(0.1\)\\
C2 feasibility threshold & \(<0.03\)\\
Literature-baseline cost weights & \([2,1,0.5,0.5]^{\mathsf T}\)\\
No-AL cost / graph weights & \([5,2,1,1]^{\mathsf T}/5\)\\
\bottomrule
\end{tabular}
\end{table}

\subsection{Comparison with Learned Baselines}
\label{subsec:families}

We implement environment-matched versions of multi-agent deep Q-networks (MADQN)~\cite{liang2019spectrum}, temporal GNN+MATD3~\cite{huang2026htdrive}, and Graph+MAPPO~\cite{singh2026gfedmap}. MADQN uses vehicle-specific Q-networks over 42 resource--power actions. GNN+MATD3 uses GraphSAGE--GRU encoding and twin centralized critics. Graph+MAPPO uses a graph actor and centralized value network. Each baseline optimizes
\begin{align}
r_t^{\mathrm{base}}
&=u_t-\bar{\bm\nu}^{\mathsf T}
[\bar{\bm c}_t-\bm\epsilon]_+,\nonumber\\
\bar{\bm c}_t&=N^{-1}\sum_n\bm c_{n,t},\qquad
\bar{\bm\nu}=[2,1,0.5,0.5]^{\mathsf T}.
\label{eq:baseline_reward}
\end{align}
All methods train for 2000 episodes and are evaluated over 100 held-out raw and filtered episodes per seed.

\begin{table*}[t]
\centering
\caption{Five-seed comparison under the mixed-constraint setting. Low-TTC denotes episode-mean failure, while C2 max is the maximum epoch-wise expected failure.}
\label{tab:families}
\footnotesize
\resizebox{\textwidth}{!}{%
\begin{tabular}{@{}lrrrrrrrrrrc@{}}
\toprule
Method & Raw utility & Raw PDR & Raw low-TTC & Raw C2 max &
Filtered utility & Filtered PDR & Filtered low-TTC & Filtered C2 max &
Filtered CBR & Resource repair & Feasible\\
\midrule
MADQN & \(53.714\pm7.120\) & \(\mathbf{0.8098}\pm0.0342\)
& \(0.0521\pm0.0159\) & \(0.1021\pm0.0179\)
& \(71.495\pm1.565\) & \(\mathbf{0.9406}\pm0.0037\)
& \(\mathbf{0.0044}\pm0.0005\) & \(0.0205\pm0.0054\)
& \(0.5004\pm0.0135\) & \(\mathbf{0.3952}\pm0.0086\) & 5/5\\
GNN+MATD3 & \(\mathbf{128.638}\pm1.680\) & \(0.5629\pm0.0698\)
& \(0.0842\pm0.0278\) & \(0.1522\pm0.0371\)
& \(\mathbf{88.039}\pm1.241\) & \(0.9292\pm0.0035\)
& \(0.0048\pm0.0012\) & \(\mathbf{0.0192}\pm0.0061\)
& \(0.4458\pm0.0093\) & \(0.4796\pm0.0250\) & 5/5\\
Graph+MAPPO & \(97.109\pm14.276\) & \(0.6207\pm0.1297\)
& \(0.0723\pm0.0422\) & \(0.1293\pm0.0526\)
& \(84.501\pm1.949\) & \(0.9332\pm0.0051\)
& \(0.0053\pm0.0008\) & \(0.0212\pm0.0055\)
& \(0.4536\pm0.0151\) & \(0.4557\pm0.0378\) & 5/5\\
\rowcolor{proposedrow}
PA-STGAT & \(105.202\pm13.633\) & \(0.7474\pm0.0545\)
& \(\mathbf{0.0312}\pm0.0120\) & \(\mathbf{0.0694}\pm0.0212\)
& \(84.015\pm1.681\) & \(0.9253\pm0.0051\)
& \(0.0057\pm0.0014\) & \(0.0224\pm0.0042\)
& \(\mathbf{0.4334}\pm0.0141\) & \(0.4238\pm0.0181\) & 5/5\\
\bottomrule
\end{tabular}}
\end{table*}

PA-STGAT improves raw PDR over GNN+MATD3 by \(0.1845\pm0.0610\) and reduces raw low-TTC failure by \(0.0530\pm0.0236\), with a utility reduction of \(23.437\pm12.646\). 
These paired intervals exclude zero.
MADQN achieves higher mean raw PDR but lower utility; its paired PDR difference from PA-STGAT is unresolved.
Under filtered execution, all three baselines have higher mean PDR and lower mean low-TTC failure and C2 max than PA-STGAT.
GNN+MATD3 achieves \(4.024\pm2.467\) higher utility, while PA-STGAT requires \(0.0558\pm0.0222\) fewer resource repairs.
All four methods meet the nominal empirical feasibility criteria, while none of the raw policies does.
The clearest reliability gains therefore concern the raw policy; filtered execution exhibits utility--reliability tradeoffs across methods.
At eight vehicles, the CBR budget is inactive.
Lower CBR indicates reduced resource occupancy, but alone does not establish improved interference coordination.

\begin{figure*}[t]
\centering
\includegraphics[width=0.95\textwidth]{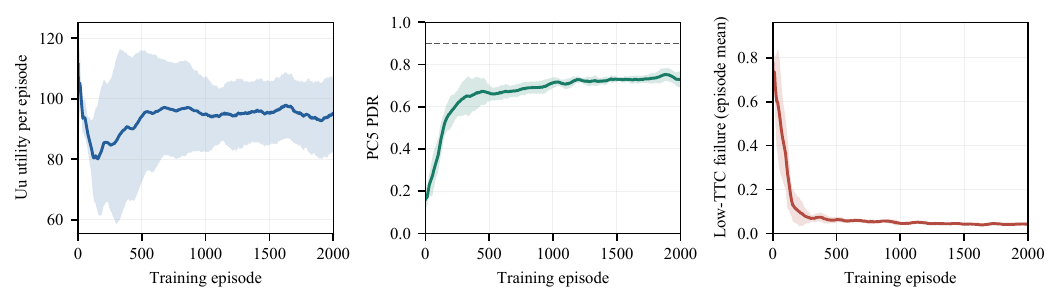}
\caption{PA-STGAT learning curves from five training seeds. Lines show trailing 100-episode means, and shaded regions show 95\% Student-\(t\) confidence intervals. Training uses decoded raw actions without the deployment filter. The dashed line marks the PDR target. Low-TTC failure is shown as an episode mean.}
\label{fig:learning_curve}
\end{figure*}

Figure~\ref{fig:learning_curve} shows the 2000-episode repair-free training process. Raw PDR increases and low-TTC failure decreases during training, while utility shows larger variation across seeds. Constraint feasibility is evaluated using the held-out C2 max in Table~\ref{tab:families}.

\begin{table*}[!t]
\centering
\caption{Architecture ablation under the mixed-constraint setting.}
\label{tab:ablation}
\footnotesize
\resizebox{\textwidth}{!}{%
\begin{tabular}{@{}lrrrrrrrrc@{}}
\toprule
Actor & Raw utility & Raw PDR & Raw low-TTC & Raw C2 max &
Filtered utility & Filtered PDR & Filtered C2 max & Filtered CBR & Feasible\\
\midrule
\rowcolor{proposedrow}
PA-STGAT & \(105.202\pm13.633\) & \(0.7474\pm0.0545\)
& \(0.0312\pm0.0120\) & \(0.0694\pm0.0212\)
& \(84.015\pm1.681\) & \(0.9253\pm0.0051\)
& \(0.0224\pm0.0042\) & \(0.4334\pm0.0141\) & 5/5\\
No risk relation & \(94.591\pm16.472\) & \(0.7967\pm0.0632\)
& \(0.0207\pm0.0089\) & \(0.0695\pm0.0197\)
& \(81.024\pm4.916\) & \(0.9260\pm0.0043\)
& \(0.0245\pm0.0056\) & \(0.4314\pm0.0066\) & 4/5\\
Snapshot GAT & \(107.931\pm20.235\) & \(0.7575\pm0.0308\)
& \(0.0236\pm0.0029\) & \(0.0624\pm0.0101\)
& \(83.506\pm3.938\) & \(0.9264\pm0.0018\)
& \(0.0241\pm0.0069\) & \(0.4374\pm0.0133\) & 4/5\\
Temporal MLP & \(54.691\pm15.143\) & \(\mathbf{0.8519}\pm0.1287\)
& \(\mathbf{0.0194}\pm0.0196\) & \(\mathbf{0.0593}\pm0.0257\)
& \(70.823\pm2.998\) & \(\mathbf{0.9480}\pm0.0026\)
& \(\mathbf{0.0194}\pm0.0061\) & \(0.5232\pm0.0095\) & 5/5\\
\bottomrule
\end{tabular}}
\end{table*}

\subsection{Architecture Ablations}
\label{subsec:ablation}

We remove the risk relation, temporal GRU, and graph aggregation separately. All variants use the same constraint update, 2000 training episodes, and 100 held-out episodes per seed.

Graph aggregation increases filtered utility over the Temporal MLP by \(13.193\pm2.667\) and reduces CBR by \(0.08985\pm0.00961\), while filtered PDR decreases by \(0.02269\pm0.00280\). The MLP also has higher mean raw PDR. Removing the risk relation or temporal recurrence produces no resolved paired raw-policy difference under the nominal setting.

\subsection{Constraint Pricing and Mechanism Ablations}
\label{subsec:al_results}

We first compare adaptive AL pricing with a matched fixed-penalty No-AL policy. Both use the same actor and 500 training episodes, and both execute raw actions during training. No-AL moves the cost and graph-conflict penalties into the reward as in~\eqref{eq:noal}. We then remove transmitter credit and dual-priced graph conflict separately. The matched AL--No-AL comparison uses 100 held-out episodes per seed, while the mechanism ablations use 30.

\begin{table}[H]
\centering
\caption{AL versus No-AL with five seeds, 500 training episodes, and 100 held-out episodes.}
\label{tab:raw_al}
\scriptsize
\resizebox{\columnwidth}{!}{%
\begin{tabular}{@{}lrrrr@{}}
\toprule
Policy & PDR & C2 max & Resource repair & Feasible\\
\midrule
AL raw & \(0.6617\pm0.0600\) & \(0.0916\pm0.0243\) & 0 & 0/5\\
No-AL raw & 0 & 1.0000 & 0 & 0/5\\
\rowcolor{proposedrow}
AL filtered & \(0.9220\pm0.0020\) & \(0.0198\pm0.0040\)
& \(0.4553\pm0.0236\) & 5/5\\
No-AL filtered & \(0.9168\pm0.0013\) & \(0.0243\pm0.0043\)
& \(0.6822\pm0.0001\) & 4/5\\
\bottomrule
\end{tabular}}
\end{table}

Adaptive pricing changes the learned raw policy substantially. Relative to No-AL, it increases raw PDR by \(0.6617\pm0.0600\) and reduces episode-mean low-TTC failure by \(0.9024\pm0.0193\). After filtering, AL increases PDR by \(0.0052\pm0.0017\) and reduces Resource repairs by \(0.2269\pm0.0236\). It satisfies the feasibility test in 5/5 seeds, compared with 4/5 for No-AL.

\begin{table}[H]
\centering
\caption{Five-seed mechanism ablation after 500 training and 30 held-out episodes.}
\label{tab:mechanism_ablation}
\scriptsize
\resizebox{\columnwidth}{!}{%
\begin{tabular}{@{}lrrrr@{}}
\toprule
Variant & Raw PDR & Raw low-TTC & Filtered C2 max & Feasible\\
\midrule
\rowcolor{proposedrow}
Full AL & \(0.6494\pm0.0700\) & \(0.0466\pm0.0196\)
& \(0.0367\pm0.0157\) & 2/5\\
No credit & \(0.6796\pm0.0633\) & \(0.0436\pm0.0125\)
& \(0.0336\pm0.0008\) & 0/5\\
No graph term & 0 & 0.9590
& \(0.0417\pm0.0103\) & 0/5\\
Generic AL & \(0.0290\pm0.0752\) & \(0.9042\pm0.1361\)
& \(0.0450\pm0.0093\) & 0/5\\
\bottomrule
\end{tabular}}
\end{table}

Without graph-conflict pricing, all five deterministic checkpoints select one resource and raw PDR falls to zero. Removing both graph conflict and transmitter credit produces the same collapse in four of five checkpoints, with mean raw PDR \(0.0290\). Relative to removing graph conflict alone, the full method increases raw PDR by \(0.6494\pm0.0700\) and reduces low-TTC failure by \(0.9124\pm0.0196\). Removing transmitter credit alone produces no resolved raw-policy difference. The graph-conflict term therefore provides the clearest coordination effect in this ablation. 
C2 max varies with the held-out sample and the number of evaluation episodes. We therefore report the 30-episode mechanism study and the 100-episode AL--No-AL comparison separately.
Their feasibility counts should not be compared directly.

\subsection{Vehicle-Count Generalization}
\label{subsec:scalability}

We keep 14 PC5 resources fixed and evaluate the eight-vehicle checkpoints without adaptation at \(N\in\{8,10,12,14,16\}\). Each method uses five policy seeds and 100 held-out episodes per seed. An independent held-out set is used for this study. MADQN uses eight vehicle-specific Q-networks and therefore cannot be directly applied to larger \(N\).

\begin{table}[t]
\centering
\caption{Zero-shot vehicle-count evaluation with 14 fixed PC5 resources. Entries are five-seed means over 100 held-out episodes per seed. A dash denotes an unsupported checkpoint.}
\label{tab:scalability}
\footnotesize
\resizebox{\columnwidth}{!}{%
\begin{tabular}{@{}llrrrrc@{}}
\toprule
Method & \(N\) & Utility & PDR & CBR & C2 max & Feasible\\
\midrule
\rowcolor{proposedrow}
PA-STGAT & 8 & 82.03 & 0.9246 & 0.4448 & 0.0268 & 3/5\\
\rowcolor{proposedrow}
 & 10 & 51.60 & 0.9157 & 0.5508 & 0.0200 & 5/5\\
\rowcolor{proposedrow}
 & 12 & 35.20 & 0.9031 & 0.6330 & 0.0247 & 4/5\\
\rowcolor{proposedrow}
 & 14 & 25.60 & 0.8934 & 0.7149 & 0.0293 & 0/5\\
\rowcolor{proposedrow}
 & 16 & 19.44 & 0.8808 & 0.7867 & 0.0419 & 0/5\\
\midrule
MADQN & 8 & 71.20 & 0.9400 & 0.5015 & 0.0205 & 5/5\\
 & 10--16 & \multicolumn{5}{c}{---}\\
\midrule
GNN+MATD3 & 8 & 85.89 & 0.9282 & 0.4542 & 0.0192 & 5/5\\
 & 10 & 54.07 & 0.9166 & 0.5572 & 0.0207 & 5/5\\
 & 12 & 37.56 & 0.9043 & 0.6442 & 0.0278 & 4/5\\
 & 14 & 27.24 & 0.8944 & 0.7218 & 0.0273 & 0/5\\
 & 16 & 20.45 & 0.8818 & 0.7955 & 0.0393 & 0/5\\
\midrule
Graph+MAPPO & 8 & 82.56 & 0.9326 & 0.4629 & 0.0195 & 5/5\\
 & 10 & 52.89 & 0.9211 & 0.5687 & 0.0200 & 5/5\\
 & 12 & 36.70 & 0.9080 & 0.6552 & 0.0277 & 1/5\\
 & 14 & 26.98 & 0.8981 & 0.7355 & 0.0297 & 0/5\\
 & 16 & 20.33 & 0.8854 & 0.8064 & 0.0384 & 0/5\\
\midrule
SB-SPS & 8 & 71.04 & 0.9293 & 0.4806 & 0.0234 & 5/5\\
 & 10 & 46.87 & 0.9147 & 0.5707 & 0.0185 & 5/5\\
 & 12 & 33.31 & 0.9046 & 0.6623 & 0.0318 & 0/5\\
 & 14 & 24.68 & 0.8919 & 0.7196 & 0.0353 & 0/5\\
 & 16 & 19.15 & 0.8791 & 0.7913 & 0.0319 & 0/5\\
\bottomrule
\end{tabular}}
\end{table}

The independent nominal evaluation finds three feasible PA-STGAT seeds, compared with five in Table~\ref{tab:families}. 
This difference shows that the empirical C2 classification depends on the held-out sample.
At \(N=8\), SB-SPS achieves higher mean PDR than PA-STGAT. PA-STGAT instead achieves higher utility and lower CBR. At \(N=10\), both methods are feasible in all seeds.
At \(N=12\), PA-STGAT and GNN+MATD3 are feasible in four seeds. Graph+MAPPO is feasible in one seed, while SB-SPS is feasible in none. GNN+MATD3 achieves higher mean utility. PA-STGAT has lower mean CBR and C2 max.
At \(N=14\), both PA-STGAT and GNN+MATD3 exceed the CBR budget and fall below the PDR target. The shared graph actor supports different vehicle counts, but feasibility decreases as the fixed resource pool becomes more crowded.

\subsection{Mobility and Beacon Stress}
\label{subsec:stress}

The nominal checkpoints are evaluated without retraining under higher speed, larger vehicle count, and delayed or missing one-hop beacons. The high-speed case uses 20--33 m/s. The combined case uses 12 vehicles, 20--33 m/s, 10-ms beacon delay, and 10\% beacon loss. Each setting uses 100 held-out episodes per seed. The local filter remains active, and proposal exchange remains reliable.

\begin{table}[t]
\centering
\caption{Zero-shot stress evaluation of PA-STGAT with the local filter. Entries are five-seed means.}
\label{tab:stress}
\scriptsize
\resizebox{\columnwidth}{!}{%
\begin{tabular}{@{}lrrrrc@{}}
\toprule
Scenario & Utility & PDR & C2 max & CBR & Feasible\\
\midrule
Nominal & 83.07 & 0.9286 & 0.0211 & 0.4392 & 5/5\\
High speed & 80.39 & 0.9310 & 0.0589 & 0.4533 & 0/5\\
12 vehicles & 35.05 & 0.8983 & 0.0210 & 0.6323 & 1/5\\
10-ms delay & 83.06 & 0.9285 & 0.0211 & 0.4395 & 5/5\\
20-ms delay & 83.02 & 0.9282 & 0.0204 & 0.4394 & 5/5\\
10\% loss & 83.09 & 0.9281 & 0.0210 & 0.4391 & 5/5\\
20\% loss & 83.17 & 0.9272 & 0.0330 & 0.4388 & 1/5\\
Combined & 34.00 & 0.9036 & 0.0846 & 0.6512 & 0/5\\
\bottomrule
\end{tabular}}
\end{table}

High-speed mobility keeps mean PDR above \(0.90\), but C2 max rises to \(0.0589\). With 20\% beacon loss, mean PDR also remains above target, while only 1/5 seeds satisfy C2. The delay and 10\% loss cases remain feasible. The combined perturbation violates both C2 and CBR. These results characterize the zero-shot robustness of the learned policy under the tested perturbations.

\subsection{Independent Filter Ablation and Search Reference}
\label{subsec:filter_oracle}

Raw and filtered policies are evaluated in independent rollouts initialized from the same scenario seeds. Filtering raises nominal PDR from \(0.7507\) to \(0.9286\) and reduces C2 max from \(0.0804\) to \(0.0211\). Mean utility decreases from \(104.10\) to \(83.07\). The filter does not recover full feasibility under high-speed mobility or 12 vehicles.

\begin{table}[t]
\centering
\caption{Independent raw and filtered rollouts of the same five PA-STGAT checkpoints over 100 held-out episodes per seed.}
\label{tab:filter_ablation}
\scriptsize
\resizebox{\columnwidth}{!}{%
\begin{tabular}{@{}llrrrrc@{}}
\toprule
Scenario & Execution & Utility & PDR & C2 max & CBR & Feasible\\
\midrule
Nominal & Raw & 104.10 & 0.7507 & 0.0804 & 0.3318 & 0/5\\
 & Filtered & 83.07 & 0.9286 & 0.0211 & 0.4392 & 5/5\\
High speed & Raw & 101.29 & 0.7297 & 0.2247 & 0.3413 & 0/5\\
 & Filtered & 80.39 & 0.9310 & 0.0589 & 0.4533 & 0/5\\
12 vehicles & Raw & 55.61 & 0.5765 & 0.1215 & 0.3887 & 0/5\\
 & Filtered & 35.05 & 0.8983 & 0.0210 & 0.6323 & 1/5\\
\bottomrule
\end{tabular}}
\end{table}

A full-information Monte Carlo (MC) search samples 256 collision-free joint actions per epoch and performs one coordinate-improvement pass. Over 30 nominal 100-epoch episodes, it reaches utility \(99.94\), PDR \(0.9521\), CBR \(0.5714\), and C2 max \(0\). It serves as a nondeployable feasible reference. A zero-sidelink-interference relaxation gives a loose Uu-utility upper bound of \(176.99\).

\section{Conclusion}
\label{sec:conclusion}
PA-STGAT combined local spatio-temporal graph perception with constrained learning for decentralized vehicle communications.
Metric-preserving transmitter credits connected individual vehicles to network-level reliability metrics. 
Dual-priced graph coordination provided a direct learning signal for categorical resource conflicts.
The ablations showed that this graph coordination was the key mechanism for preventing resource-selection collapse.
PA-STGAT achieved higher mean raw reliability than the graph-based baselines. 
It also required less resource repair than GNN+MATD3 after filtering, although the filtered metrics showed a utility--reliability tradeoff.
Under the stated assumptions, the theoretical analysis established asymptotic feasibility with a utility gap bounded by the policy-solver error. 
The independent evaluations also showed that C2 feasibility near the threshold depended on the held-out sample. 
Feasibility further decreased at higher vehicle densities and under faster mobility.

\section*{Appendix}
\label{app:sec:analysis}

This appendix establishes feasibility and near-optimality of the augmented-Lagrangian formulation. The analysis follows the
actor-training structure of PA-STGAT: the multipliers and penalty coefficient
remain fixed while the actor is trained, and are updated between training
stages. A uniform solver accuracy determines the final reward gap.

\subsection*{A. Definitions and Constrained Formulation}

Fix a horizon $T<\infty$, a discount factor $\gamma\in(0,1]$, the
initial-state distribution, and the transition law. Let $s_t$ denote the
joint network state, augmented by time and the history needed for policy
evaluation, and let $a_t$ denote the (joint resource and power) action.
The shared PA-STGAT actor induces a joint policy $\pi_\theta$ through its
vehicle-local decisions. Write $\Pi$ for the declared class of admissible
joint policies; all expectations include policy and environment randomness.

Let $r_t=r(s_t,a_t)$ be the utility, denoted by $u_t$ in the main text.
Let $c_t=c(s_t,a_t)\in\mathbb R^{m_c}$ collect the signed cumulative
residuals, and let $g_t=g(s_t,a_t)\in\mathbb R^{m_g}$ collect the signed instantaneous residuals. 
Thus the constraint assigned to instantaneous enforcement enters $g_t$;
the remaining cumulative constraints enter $c_t$.
All vector inequalities and $[\cdot ]_+$ operators are componentwise. Define
\begin{align}
 J(\pi)&=\mathbb E_\pi\!\left[\sum_{t=0}^{T-1}\gamma^t r_t\right],
 \label{app:eq:reward}\\
 C(\pi)&=\frac1T\mathbb E_\pi\!\left[\sum_{t=0}^{T-1}c_t\right],
 \qquad G_t(\pi)=\mathbb E_\pi[g_t].
 \label{app:eq:costs}
\end{align}
The constrained reward optimum is
\begin{equation}
\begin{aligned}
 J^*={}&\max_{\pi\in\Pi} J(\pi)\\
 \text{subject to }&C(\pi)\leq0,\quad
 G_t(\pi)\leq0,\quad t=0,\ldots,T-1.
\end{aligned}
\label{app:eq:problem}
\end{equation}
Note that the reward is discounted; the cumulative costs are undiscounted episode
averages.
For $[x]_+=\max\{x,0\}$, set $h_t=[g_t]_+$ and define
\begin{align}
 H(\pi)&=\mathbb E_\pi\!\left[\sum_{t=0}^{T-1}\gamma^t h_t\right],  \quad
 Q(\pi)=\mathbb E_\pi\!\left[\sum_{t=0}^{T-1}
                    \gamma^t\|h_t\|_2^2\right].
 \label{app:eq:Q}
\end{align}
Let $d=m_c+m_g$ and
$\lambda=(\mu^\mathsf T,\nu^\mathsf T)^\mathsf T\in\mathbb R_+^d$ as the Lagrangian multipliers,
where $\mu$ and $\nu$ price cumulative and instantaneous residuals,
respectively. For $\rho>0$, the training objective is
\begin{equation}
 R_\rho(\pi,\lambda)
 =J(\pi)-\mu^\mathsf T C(\pi)-\nu^\mathsf T H(\pi)
       -\frac{\rho}{2}Q(\pi).
\label{app:eq:surrogate}
\end{equation}
The cumulative residuals enter linearly. Quadratic penalties apply to the
positive instantaneous residuals inside the expected time sum.
For one constraint of each type, these definitions reduce to the scalar
notation $\lambda=(\mu,\nu)$.

For fixed $(\lambda_k,\rho_k)$, the per-step training reward is
\begin{equation}
 \widetilde r_{k,t}
 =r_t -\nu_k^\mathsf T h_t-\frac{\rho_k}{2}\|h_t\|_2^2 -\frac{\mu_k^\mathsf T c_t}{T\gamma^t}.
\label{app:eq:shaped}
\end{equation}
It satisfies
$\mathbb E_\pi[\sum_{t=0}^{T-1}\gamma^t\widetilde r_{k,t}]
=R_{\rho_k}(\pi,\lambda_k)$.
The factor $\gamma^{-t}$ preserves the undiscounted cumulative budgets. To learn this in practice, one can input time $t$ into the network together with $s_t,a_t$. 

\subsection*{B. Assumptions and Iterative Updates}

\noindent\textbf{Assumption 1 (bounded and closed performance).}
The functions $r,c,g$ are bounded, and the attainable performance set $\mathcal V=\{(J(\pi),C(\pi),H(\pi),Q(\pi)):\pi\in\Pi\}$
is compact. The class $\Pi$ is closed under finite episode-level mixtures:
one policy is selected at the start of an episode and executed throughout
that episode. 

\medskip
\noindent\textbf{Assumption 2 (instantaneously feasible optimum).}
Problem~\eqref{app:eq:problem} admits an optimizer $\pi^*$ satisfying
$H(\pi^*)=0$. Equivalently, $g_t\leq0$ almost surely under $\pi^*$ at
every $t<T$. This is the finite-horizon counterpart of the pointwise
optimal-policy condition used in the augmented-Lagrangian analysis of
Li et al.~\cite[Proposition~3]{li2021alm}.

% Since all $\gamma^t$ are positive, this condition gives $Q(\pi^*)=0$ and preserves the original constrained reward benchmark when instantaneous residuals are clipped.

\medskip
\noindent\textbf{Assumption 3 ($\delta$-capable policy solver).}
For some constant $\delta\geq0$, every training stage returns
$\pi_k\in\Pi$ such that
\begin{equation}
 R_{\rho_k}(\pi_k,\lambda_k)
 \geq\max_{\pi\in\Pi}R_{\rho_k}(\pi,\lambda_k)-\delta.
\label{app:eq:capability}
\end{equation}
The same $\delta$ applies to every iteration $k$. In PA-STGAT,
$\pi_k=\pi_{\theta_k}$, and this condition measures the solver's accuracy
in augmented-objective value. Actor approximation, critic estimation, and
the effect of the graph-conflict training term are included in this
accuracy requirement for~\eqref{app:eq:surrogate}.

\medskip
Initialize $\lambda_0=0$ and $\rho_0>0$. Choose step sizes $\eta_k$ and
multiplier bounds $M_k$ satisfying
\begin{gather}
 0<\eta_{k+1}\leq\eta_k,\qquad \eta_k\longrightarrow0,
 \label{app:eq:eta}\\
 \begin{gathered}
 0<M_k\leq M_{k+1},\qquad M_k\longrightarrow\infty,\\
 \frac{M_k^2}{(k+1)\eta_k}\longrightarrow0.
 \end{gathered}
 \label{app:eq:bounds}
\end{gather}
At each iteration $k=0,1,\ldots$, perform two steps:
\begin{enumerate}
 \item Hold $(\lambda_k,\rho_k)$ fixed and find $\pi_k$ that meets
       \eqref{app:eq:capability}.
 \item Evaluate the returned policy and update
 \begin{align}
  q_k&=(C(\pi_k)^\mathsf T,H(\pi_k)^\mathsf T)^\mathsf T,
       \label{app:eq:residual}\\
  \lambda_{k+1}
     &=\operatorname{Proj}_{[0,M_{k+1}]^d}
                  (\lambda_k+\eta_k q_k),
       \label{app:eq:dual}\\
  \rho_{k} & \geq 0.
       \label{app:eq:penalty}
 \end{align}
\end{enumerate}
% Here $\operatorname{Proj}$ is Euclidean projection. 
Both updated parameters enter the next policy-training stage $k+1$. 
The analysis uses exact expected episode residuals $q_k$. That could be extended to noisy, estimated $\hat{q}_k$ cases, which would lead to a probabilistic bound. However, that may introduce much more difficulties without providing more insights. Thus we keep with the $q_k$ case. Note that the norm of $q_k$ can be uniformly bounded by some $B < \infty$.

% After $K$ iterations, define $\bar\pi_K$ by selecting one of
% $\pi_0,\ldots,\pi_{K-1}$ uniformly at episode start. In the multi-vehicle
% system, all vehicles use the same selected actor index. Consequently,
% \begin{equation}
%  F(\bar\pi_K)=\frac1K\sum_{k=0}^{K-1}F(\pi_k),
%  \qquad F\in\{J,C,H,Q\}.
% \label{app:eq:mixture}
% \end{equation}
% This is a mixture of complete episode policies.

% Assumption~1 gives a finite constant
% $B=\sup_{\pi\in\Pi}\|(C(\pi)^\mathsf T,H(\pi)^\mathsf T)^\mathsf T\|_2$.
% Define the outer-loop error
% \begin{equation}
%  \varepsilon_K
%  =\frac{dM_{K-1}^2}{2K\eta_{K-1}}
%    +\frac{B^2}{2K}\sum_{k=0}^{K-1}\eta_k.
% \label{app:eq:outer-error}
% \end{equation}
% Conditions~\eqref{app:eq:eta}--\eqref{app:eq:bounds} imply
% $\varepsilon_K\to0$.

\subsection*{C. Feasibility and Near-Optimality}

\begin{theorem}[Uniform-accuracy iteration guarantee]
\label{app:thm:main}
Under Assumptions~1--3 and
\eqref{app:eq:eta}--\eqref{app:eq:penalty}, the uniform episode mixture $F(\bar\pi_K) \triangleq \frac{1}{K} \sum_{k=0}^{K-1}F(\pi_k)$, where $F \in \{J,C, H,Q \}$, satisfies, for every $K $, 
\begin{equation}
 J(\bar\pi_K)\geq J^*-\delta-\varepsilon_K,
\label{app:eq:finite-reward}
\end{equation}
where $\varepsilon_K
 =\frac{dM_{K-1}^2}{2K\eta_{K-1}}
   +\frac{B^2}{2K}\sum_{k=0}^{K-1}\eta_k.$
For the fixed horizon $T$, as $K\to\infty$, $ \varepsilon_K \to 0$. Moreover, 
\begin{gather}
 \begin{gathered}
 [C(\bar\pi_K)]_+\longrightarrow0,\\
 H(\bar\pi_K)\longrightarrow0,\qquad Q(\bar\pi_K)\longrightarrow0,
 \end{gathered}
 \label{app:eq:feasibility}\\
 [G_t(\bar\pi_K)]_+\longrightarrow0,
 \qquad t=0,\ldots,T-1,
 \label{app:eq:instantaneous}\\
 J^*-\delta\leq\liminf_{K\to\infty}J(\bar\pi_K)
 \leq\limsup_{K\to\infty}J(\bar\pi_K)\leq J^*.
 \label{app:eq:reward-limit}
\end{gather}
Thus a $\delta$-capable solver achieves asymptotic feasibility with a
reward gap of at most $\delta$. In particular, $\delta=0$ gives
$J(\bar\pi_K)\to J^*$.
\end{theorem}

\begin{IEEEproof}
All norms below are Euclidean. We establish the result in four steps.

\emph{Step 1: Comparison with an optimal feasible policy.}
Assumption~2 gives $C(\pi^*)\leq0$ and
$H(\pi^*)=0$, $Q(\pi^*)=0$. Hence Assumption~3 and $\mu_k\geq0$ imply
\begin{align}
 R_{\rho_k}(\pi_k,\lambda_k)
 &\geq R_{\rho_k}(\pi^*,\lambda_k)-\delta\nonumber\\
 &=J^*-\mu_k^\mathsf T C(\pi^*)-\delta
 \geq J^*-\delta.
\label{app:eq:comparison}
\end{align}

\emph{Step 2: Average control of the multiplier updates.}
Fix any $\lambda\in\mathbb R_+^d$. Since $M_k\to\infty$, there is a
fixed index $m$ such that $\lambda\in[0,M_m]^d$.
For $k\geq m$, nonexpansiveness of projection in~\eqref{app:eq:dual}
and $\|q_k\|\leq B$ give
\begin{equation}
 \begin{aligned}
 \|\lambda_{k+1}-\lambda\|^2
 \leq{}&\|\lambda_k-\lambda\|^2 +2\eta_k(\lambda_k-\lambda)^\mathsf T q_k+\eta_k^2B^2.
 \end{aligned}
\label{app:eq:projection}
\end{equation}
Put $D_k=\|\lambda_k-\lambda\|^2$. Rearranging yields
\begin{equation}
 (\lambda-\lambda_k)^\mathsf Tq_k
 \leq\frac{D_k-D_{k+1}}{2\eta_k}+\frac{B^2\eta_k}{2}.
\label{app:eq:one-step}
\end{equation}
For $K>m$, summation by parts gives
\begin{align}
 \sum_{k=m}^{K-1}\frac{D_k-D_{k+1}}{2\eta_k}
 ={}&\frac{D_m}{2\eta_m}-\frac{D_K}{2\eta_{K-1}}\nonumber\\
 &+\frac12\sum_{k=m+1}^{K-1}
 \left(\frac1{\eta_k}-\frac1{\eta_{k-1}}\right)D_k.
\label{app:eq:summation}
\end{align}
For $m\leq k<K$, both $\lambda_k$ and $\lambda$ belong to
$[0,M_{K-1}]^d$, so $D_k\leq dM_{K-1}^2$.
The coefficients in the last sum are nonnegative because $\eta_k$ is
nonincreasing. Dropping the nonpositive term containing $D_K$ and
telescoping these coefficients bounds~\eqref{app:eq:summation} by
$dM_{K-1}^2/(2\eta_{K-1})$.
Summing~\eqref{app:eq:one-step} and dividing by $K$ therefore gives
\begin{equation}
 \frac1K\sum_{k=m}^{K-1}(\lambda-\lambda_k)^\mathsf Tq_k
 \leq\varepsilon_K\longrightarrow0.
\label{app:eq:average-control}
\end{equation}

\emph{Step 3: Transfer to the uniform episode mixture.}
By~\eqref{app:eq:comparison},
\begin{align}
 J(\pi_k)-\lambda^\mathsf Tq_k
 &\geq R_{\rho_k}(\pi_k,\lambda_k)
       -(\lambda-\lambda_k)^\mathsf Tq_k\nonumber\\
 &\geq J^*-\delta-(\lambda-\lambda_k)^\mathsf Tq_k,
\label{app:eq:transfer}
\end{align}
where the first inequality uses $\rho_kQ(\pi_k)/2\geq0$.
Average over $k=m,\ldots,K-1$ and apply~\eqref{app:eq:average-control}.
The omitted prefix has a fixed finite length, so its contribution divided
by $K$ vanishes.  Via the definition of $\bar{\pi}_K$, for every fixed
$\mu,\nu\geq0$,
\begin{equation}
 \begin{gathered}
 \liminf_{K\to\infty}\bigl\{
 J(\bar\pi_K)-\mu^\mathsf T C(\bar\pi_K)
                  -\nu^\mathsf T H(\bar\pi_K)\bigr\} \geq J^*-\delta.
 \end{gathered}
\label{app:eq:all-comparators}
\end{equation}
For $\lambda=0$, choose $m=0$; the same calculation before taking limits
proves~\eqref{app:eq:finite-reward} for every $K\geq1$.

\emph{Step 4: Feasibility and the reward gap.}
Take any convergent subsequence of the performance vectors of
$\bar\pi_K$. Mixture closure and compactness in Assumption~1 ensure
that its limit is the performance vector of some $\widehat\pi\in\Pi$.
Passing to this limit in~\eqref{app:eq:all-comparators} gives
\begin{equation}
 \begin{gathered}
 J(\widehat\pi)-\mu^\mathsf T C(\widehat\pi)
                  -\nu^\mathsf T H(\widehat\pi)
 \geq J^*-\delta, \quad \forall \mu,\nu\geq0.
 \end{gathered}
\label{app:eq:limit-comparison}
\end{equation}
If a component of $C(\widehat\pi)$ were positive, increasing only its
multiplier would contradict~\eqref{app:eq:limit-comparison}.
The same argument for $H(\widehat\pi)\geq0$ shows
$H(\widehat\pi)=0$. Thus $C(\widehat\pi)\leq0$ and
$g_t\leq0$ almost surely at every $t<T$; in particular,
$Q(\widehat\pi)=0$ and $\widehat\pi$ is feasible
for~\eqref{app:eq:problem}.
Feasibility gives $J(\widehat\pi)\leq J^*$, while setting
$\mu=\nu=0$ in~\eqref{app:eq:limit-comparison} gives
$J(\widehat\pi)\geq J^*-\delta$.

Every accumulation vector therefore has nonpositive cumulative
residuals, zero $H$ and $Q$, and reward in $[J^*-\delta,J^*]$.
Compactness now proves~\eqref{app:eq:feasibility} and
\eqref{app:eq:reward-limit}.
Finally, for every instantaneous component $i$ and time $t<T$,
\begin{equation*}
 0\leq[G_t(\bar\pi_K)_i]_+
 \leq\mathbb E_{\bar\pi_K}\!\left[[(g_t)_i]_+\right]
 \leq\frac{H(\bar\pi_K)_i}{\gamma^t}\longrightarrow0,
\end{equation*}
which establishes~\eqref{app:eq:instantaneous}.
\end{IEEEproof}

\begingroup
\providecommand{\BIBdecl}{\fontsize{7.2}{8.0}\selectfont}
\bibliographystyle{IEEEtran}
\bibliography{references}

@article{dai2026agenticcaching,
  author  = {Y. Dai and H. Yang and M. Zhou and W. Sun and Y. Zhang},
  title   = {{Agentic AI for Low-Latency Edge Caching and Resource Allocation in Internet of Vehicles}},
  journal = {{IEEE Trans. Cogn. Commun. Netw.}},
  volume  = {12},
  pages   = {12201--12210},
  year    = {2026},
  doi     = {10.1109/TCCN.2026.3727467}
}

@techreport{3gpp22186,
  author       = {{3GPP}},
  title        = {{Service requirements for enhanced V2X scenarios}},
  institution  = {{3rd Generation Partnership Project (3GPP)}},
  type         = {{Technical Specification}},
  number       = {{TS 22.186}},
  revision     = {{V19.0.0}},
  month        = oct,
  year         = {2025},
  note         = {{Release 19}}
}

@techreport{3gpp23287,
  author       = {{3GPP}},
  title        = {{Architecture enhancements for 5G System (5GS) to support Vehicle-to-Everything (V2X) services}},
  institution  = {{3rd Generation Partnership Project (3GPP)}},
  type         = {{Technical Specification}},
  number       = {{TS 23.287}},
  revision     = {{V19.0.0}},
  month        = oct,
  year         = {2025},
  note         = {{Release 19}}
}

@techreport{3gpp38300,
  author       = {{3GPP}},
  title        = {{NR and NG-RAN Overall description; Stage-2}},
  institution  = {{3rd Generation Partnership Project (3GPP)}},
  type         = {{Technical Specification}},
  number       = {{TS 38.300}},
  revision     = {{V19.3.0}},
  month        = jun,
  year         = {2026},
  note         = {{Release 19}}
}

@techreport{3gpp38214,
  author       = {{3GPP}},
  title        = {{NR; Physical layer procedures for data}},
  institution  = {{3rd Generation Partnership Project (3GPP)}},
  type         = {{Technical Specification}},
  number       = {{TS 38.214}},
  revision     = {{V19.2.0}},
  month        = feb,
  year         = {2026},
  note         = {{Release 19}}
}

@techreport{3gpp38321,
  author       = {{3GPP}},
  title        = {{NR; Medium Access Control (MAC) protocol specification}},
  institution  = {{3rd Generation Partnership Project (3GPP)}},
  type         = {{Technical Specification}},
  number       = {{TS 38.321}},
  revision     = {{V19.1.0}},
  month        = feb,
  year         = {2026},
  note         = {{Release 19}}
}

@techreport{3gpp37885,
  author       = {{3GPP}},
  title        = {{Study on evaluation methodology of new Vehicle-to-Everything (V2X) use cases for LTE and NR}},
  institution  = {{3rd Generation Partnership Project (3GPP)}},
  type         = {{Technical Report}},
  number       = {{TR 37.885}},
  revision     = {{V15.3.0}},
  month        = jun,
  year         = {2019},
  note         = {{Release 15}}
}

@techreport{3gpp38901,
  author       = {{3GPP}},
  title        = {{Study on channel model for frequencies from 0.5 to 100 GHz}},
  institution  = {{3rd Generation Partnership Project (3GPP)}},
  type         = {{Technical Report}},
  number       = {{TR 38.901}},
  revision     = {{V19.2.0}},
  month        = jan,
  year         = {2026},
  note         = {{Release 19}}
}

@inproceedings{reyhanoglu2023qos,
  author    = {A. Reyhanoglu and others},
  title     = {{Machine Learning Aided NR-V2X Quality of Service Predictions}},
  booktitle = {{Proc. IEEE Veh. Netw. Conf. (VNC)}},
  address   = {Istanbul, Turkiye},
  pages     = {183--186},
  year      = {2023},
  doi       = {10.1109/VNC57357.2023.10136346}
}

@article{nguyen2023shadowing,
  author  = {H. T. Nguyen and M. Noor-A-Rahim and Y. L. Guan and D. Pesch},
  title   = {{Cellular V2X Communications in the Presence of Big Vehicle Shadowing: Performance Analysis and Mitigation}},
  journal = {{IEEE Trans. Veh. Technol.}},
  volume  = {72},
  number  = {3},
  pages   = {3764--3776},
  month   = mar,
  year    = {2023},
  doi     = {10.1109/TVT.2022.3212704}
}

@article{mondal2026fusion,
  author  = {S. Mondal},
  title   = {{Three-Layer V2X Fusion With PC5/Uu Dual-Interface Routing for Real-Time Cooperative Perception in Autonomous Driving}},
  journal = {{IEEE Trans. Veh. Technol.}},
  year    = {2026},
  note    = {{Early Access}, doi: 10.1109/TVT.2026.3708424}
}

@article{lin2026riskaware,
  author  = {Z. Lin and M. Zheng and Z. Huang and S. Zhou},
  title   = {{Risk-Aware Joint Communication and Control Resource Allocation for Secure Vehicle Platoons}},
  journal = {{IEEE Wireless Commun. Lett.}},
  volume  = {15},
  pages   = {835--839},
  year    = {2026},
  doi     = {10.1109/LWC.2025.3637860}
}

@article{petho2024cyberrisk,
  author  = {Z. Peth{\H{o}} and T. M. Kaz{\'a}r and Z. Szalay and {\'A}. T{\"o}r{\"o}k},
  title   = {{Quantifying Cyber Risks: The Impact of DoS Attacks on Vehicle Safety in V2X Networks}},
  journal = {{IEEE Trans. Intell. Transp. Syst.}},
  volume  = {25},
  number  = {11},
  pages   = {18591--18600},
  month   = nov,
  year    = {2024},
  doi     = {10.1109/TITS.2024.3436840}
}

@inproceedings{onsu2026semantic,
  author    = {M. A. Onsu and P. Lohan and B. Kantarci and A. Syed and M. Andrews and S. Kennedy},
  title     = {{Spatiotemporal Semantic V2X Framework for Cooperative Collision Prediction}},
  booktitle = {{Proc. IEEE Int. Conf. Commun. (ICC)}},
  address   = {Glasgow, U.K.},
  pages     = {1--6},
  year      = {2026},
  doi       = {10.1109/ICC59461.2026.11587981}
}

@article{wang2025pcars,
  author  = {B. Wang and J. Zheng and N. Mitton and C. Li},
  title   = {{A Packet Collision Avoidance Resource Selection Scheme for Reliable Intra-Platoon Message Delivery in a C-V2X Network}},
  journal = {{IEEE Trans. Veh. Technol.}},
  volume  = {74},
  number  = {7},
  pages   = {11366--11377},
  month   = jul,
  year    = {2025},
  doi     = {10.1109/TVT.2025.3544826}
}

@article{zhou2025resource,
  author  = {X. Zhou and F. Hui and J. Liu and W. Wang and J. Zhang},
  title   = {{Resource Reservation in C-V2X Networks for Dynamic Traffic Environments: From Vehicle Density-Driven to Deep Reinforcement Learning}},
  journal = {{IEEE Trans. Veh. Technol.}},
  volume  = {74},
  number  = {11},
  pages   = {17429--17444},
  month   = nov,
  year    = {2025},
  doi     = {10.1109/TVT.2025.3578083}
}

@article{saad2025federatednrmode2,
  author  = {M. M. Saad and M. A. Tariq and M. Ajmal and D. Kim and G. Srivastava},
  title   = {{Federated Multiagent Reinforcement Learning for Resource Allocation in NR-V2X Mode 2}},
  journal = {{IEEE Internet Things J.}},
  volume  = {12},
  number  = {13},
  pages   = {23402--23417},
  month   = jul,
  year    = {2025},
  doi     = {10.1109/JIOT.2025.3555195}
}

@article{liang2019spectrum,
  author  = {L. Liang and H. Ye and G. Y. Li},
  title   = {{Spectrum Sharing in Vehicular Networks Based on Multi-Agent Reinforcement Learning}},
  journal = {{IEEE J. Sel. Areas Commun.}},
  volume  = {37},
  number  = {10},
  pages   = {2282--2292},
  month   = oct,
  year    = {2019},
  doi     = {10.1109/JSAC.2019.2933962}
}

@article{singh2026gfedmap,
  author  = {P. Singh and B. Hazarika and W.-J. Huang},
  title   = {{Graph-Based Federated Multiagent DRL for Semantic and Intent-Aware V2X Communication}},
  journal = {{IEEE Internet Things J.}},
  volume  = {13},
  number  = {5},
  pages   = {8143--8160},
  month   = mar,
  year    = {2026},
  doi     = {10.1109/JIOT.2025.3616381}
}

@article{huang2026htdrive,
  author  = {Y. Huang and F. Han and S. Zhao},
  title   = {{HT-DRIVE: Heterogeneous-Temporal GNN-Aided Multi-Agent Reinforcement Learning for Resource Allocation in Platoon-Based V2X Networks}},
  journal = {{IEEE Trans. Commun.}},
  volume  = {74},
  pages   = {6541--6557},
  year    = {2026},
  doi     = {10.1109/TCOMM.2026.3678434}
}

@article{zheng2026aoiplatoon,
  author  = {Y. Zheng and L. D. Nguyen and T. Q. Duong},
  title   = {{Multiagent Deep Reinforcement Learning for Optimal Resource Allocation in AoI-Aware Energy-Efficient Platoon-Based C-V2X Networks}},
  journal = {{IEEE Internet Things J.}},
  volume  = {13},
  number  = {6},
  pages   = {10814--10828},
  month   = mar,
  year    = {2026},
  doi     = {10.1109/JIOT.2025.3650400}
}

@article{wang2026multiagentdrlv2xresource,
  author  = {S. Wang and L. Lei and P. Maheshwari and S. Bellefeuille and K. Zheng},
  title   = {{Multi-Agent Reinforcement Learning for V2X Resource Allocation: Disentangling MARL Challenges Through Benchmarking}},
  journal = {{arXiv preprint arXiv:2603.06607}},
  year    = {2026}
}

@article{chen2023gatstc,
  author  = {Y. Chen and others},
  title   = {{Graph Attention Network With Spatial-Temporal Clustering for Traffic Flow Forecasting in Intelligent Transportation System}},
  journal = {{IEEE Trans. Intell. Transp. Syst.}},
  volume  = {24},
  number  = {8},
  pages   = {8727--8737},
  month   = aug,
  year    = {2023},
  doi     = {10.1109/TITS.2022.3208952}
}

@article{singh2023learningcmdp,
  author  = {R. Singh and A. Gupta and N. B. Shroff},
  title   = {{Learning in Constrained Markov Decision Processes}},
  journal = {{IEEE Trans. Control Netw. Syst.}},
  volume  = {10},
  number  = {1},
  pages   = {441--453},
  month   = mar,
  year    = {2023},
  doi     = {10.1109/TCNS.2022.3203361}
}

@inproceedings{li2021alm,
  author    = {J. Li and D. Fridovich-Keil and S. Sojoudi and C. J. Tomlin},
  title     = {{Augmented Lagrangian Method for Instantaneously Constrained Reinforcement Learning Problems}},
  booktitle = {{Proc. IEEE Conf. Decis. Control (CDC)}},
  pages     = {2982--2989},
  year      = {2021},
  doi       = {10.1109/CDC45484.2021.9683088}
}

@article{zhu2026maddpg,
  author  = {L. Zhu and K. Yang and J. Li and Y. Yang and Q. Yang and X. Gao},
  title   = {{MADDPG Based Path Planning for Multi-UAV Information Collection in IoT Networks}},
  journal = {{IEEE Trans. Netw. Sci. Eng.}},
  volume  = {13},
  pages   = {9300--9319},
  year    = {2026},
  doi     = {10.1109/TNSE.2026.3690003}
}

@ARTICLE{11673126,
  author={Wu, Qiong and Zhang, Wenjun and Fan, Pingyi and Wang, Kezhi and Fan, Qiang and Chen, Wen and Mao, Guoqiang and Letaief, Khaled B.},
  journal={IEEE Commun. Surv. Tutor.}, 
  title={From {LLMs} to Agentic and Embodied {AI} for Next-Generation Intelligent Vehicular Communications: A Comprehensive Survey}, 
  year={2026},
  volume={28},
  number={1},
  pages={6983--7020},
  month={Sep.},}

@ARTICLE{11667010,
  author={Dai, Yueyue and Yang, Huijiong and Zhou, Man and Sun, Wen and Zhang, Yan},
  journal={IEEE Trans. Cogn. Commun. Netw.}, 
  title={Agentic {AI} for Low-Latency Edge Caching and Resource Allocation in Internet of Vehicles}, 
  year={2026},
  volume={12},
  number={1},
  pages={12201--12210},
  month={Aug.},}

@ARTICLE{10701015,
  author={Camajori Tedeschini, Bernardo and Brambilla, Mattia and Nicoli, Monica and Win, Moe Z.},
  journal={IEEE Trans. Intell. Veh.}, 
  title={Multi-Agent Reinforcement Learning for Distributed Cooperative Vehicular Positioning}, 
  year={2025},
  volume={10},
  number={7},
  pages={4052--4067},
  month={Jul.}}

@ARTICLE{10697115,
  author={Ji, Maoxin and Wu, Qiong and Fan, Pingyi and Cheng, Nan and Chen, Wen and Wang, Jiangzhou and Letaief, Khaled B.},
  journal={IEEE Int. Things J.}, 
  title={Graph Neural Networks and Deep Reinforcement Learning-Based Resource Allocation for {V2X} Communications}, 
  year={2025},
  volume={12},
  number={4},
  pages={3613--3628},
  month={Feb.}}

@ARTICLE{10632058,
  author={Wang, Zixin and Bennis, Mehdi and Zhou, Yong},
  journal={IEEE Trans. Wireless Commun.}, 
  title={Graph Attention-Based MADRL for Access Control and Resource Allocation in Wireless Networked Control Systems}, 
  year={2024},
  volume={23},
  number={11},
  pages={16076--16090},
  month={Nov.}}

@ARTICLE{10431724,
  author={Hu, Pihe and Chen, Yu and Pan, Ling and Fang, Zhixuan and Xiao, Fu and Huang, Longbo},
  journal={IEEE/ACM Trans. Netw.}, 
  title={Multi-User Delay-Constrained Scheduling With Deep Recurrent Reinforcement Learning}, 
  year={2024},
  volume={32},
  number={3},
  pages={2344--2359},
  month={Jun.}}

@ARTICLE{10907776,
  author={Liu, Wanting and Zhang, Chengwei and Fang, Wanqing and Zhou, Kailing and Li, Yihong and Zhan, Furui and Wang, Qi and Xue, Wanli and Chen, Rong},
  journal={IEEE Trans. Intell. Transp. Syst.}, 
  title={Vehicle-Level Fairness-Oriented Constrained Multi-Agent Reinforcement Learning for Adaptive Traffic Signal Control}, 
  year={2025},
  volume={26},
  number={4},
  pages={4878--4890},
  month={Apr.}}
\endgroup

\end{document}